\documentclass[12pt]{article}
\usepackage{epsfig}
\def\apm#1{\hbox{$\pm #1$}}
\def\epm#1#2{\hbox{${\lower1pt\hbox{$\scriptstyle +#1$}}
\atop {\raise1pt\hbox{$\scriptstyle -#2$}}$}}

\def\gsim{\mathrel{\rlap{\lower4pt\hbox{\hskip1pt$\sim$}}
    \raise1pt\hbox{$>$}}}         

\def\etal{{\it et al.}}

\def\frac#1#2{{{#1}\over {#2}}}

\def\smallfrac#1#2{\hbox{${{#1}\over {#2}}$}}

\def\GeV{{\rm GeV}}

\def\bq{\bar{q}}
\catcode`@=11 
\renewcommand\section{\@startsection {section}{1}{\z@}
    {-3.5ex plus -1ex minus -.2ex}{2.3ex plus .2ex}{\bf}}
\renewcommand\subsection{\@startsection {subsection}{1}{\z@}
    {-3.5ex plus -1ex minus -.2ex}{2.3ex plus .2ex}{\it}}
\def\slash#1{\mathord{\mathpalette\c@ncel#1}}
 \def\c@ncel#1#2{\ooalign{$\hfil#1\mkern1mu/\hfil$\crcr$#1#2$}}
\def\lsim{\mathrel{\mathpalette\@versim<}}
\def\gsim{\mathrel{\mathpalette\@versim>}}
 \def\@versim#1#2{\lower0.2ex\vbox{\baselineskip\z@skip\lineskip\z@skip
       \lineskiplimit\z@\ialign{$\m@th#1\hfil##$\crcr#2\crcr\sim\crcr}}}
\catcode`@=12 

\def\PR{{\it Phys.~Rev.~}}
\def\PRL{{\it Phys.~Rev.~Lett.~}}
\def\NP{{\it Nucl.~Phys.~}}
\def\NPBPS{{\it Nucl.~Phys.~B (Proc.~Suppl.)~}}
\def\PL{{\it Phys.~Lett.~}}

\def\ZP{{\it Zeit.~Phys.~}}

\def\APP{{\it Acta.~Phys.~Pol.~}}
\def\vol#1{{\bf #1}}\def\vyp#1#2#3{\vol{#1} (#2) #3}

\def\be{\begin{equation}}
\def\ee{\end{equation}}
\def\bea{\begin{eqnarray}}
\def\eea{\end{eqnarray}}
\def\beq{\begin{equation}}
\def\eeq{\end{equation}}
\def\bq{\begin{quote}}
\def\eq{\end{quote}}
\def\gappeq{\mathrel{\rlap {\raise.5ex\hbox{$>$}} {\lower.5ex\hbox{$\sim$}}}}
\def\lappeq{\mathrel{\rlap{\raise.5ex\hbox{$<$}} {\lower.5ex\hbox{$\sim$}}}}
  \newcommand{\ccaption}[2]{
    \begin{center}
    \parbox{0.85\textwidth}{
      \caption[#1]{\small{\it{#2}}}
      }
    \end{center}
    }
\begin{document}
\parskip 0.3cm

\pagestyle{empty}
\begin{flushright}
   \begin{minipage}{4cm}
        CERN-TH/98-61   \hfill\\
        DFTT 10/98      \hfill\\
        Edinburgh 98/4  \hfill\\
        GEF-TH-5/1998   \hfill\\
        UB-ECM-PF 98/07 \hfill\\
        hep-ph/9803237  \hfill\\
   \end{minipage}
\end{flushright}
\begin{center} {\large
\bf Theoretical Analysis of Polarized Structure Functions}\\
\vspace*{0.8cm} {\bf Guido Altarelli} \\
Theoretical Physics Division, CERN, CH--1211 Geneva 23, Switzerland \\
{\it and} Universit\`a di Roma Tre, Rome, Italy \\
\vspace{0.2cm}{\bf
Richard D. Ball\footnote[1]{Royal Society University Research Fellow}} \\
Department of Physics and Astronomy, University of Edinburgh,\\
Mayfield Road, Edinburgh EH9 3JZ, Scotland\\
\vspace{0.2cm}{\bf Stefano Forte} \\
INFN, Sezione di Torino, Via P. Giuria 1, I-10125 Torino, Italy\\
{\it and} Departament ECM, Universitat de Barcelona,
Diagonal 647, E-08028 Barcelona, 
Spain\footnote[2]{IBERDROLA visiting professor}
\vspace{0.2cm}{\bf Giovanni Ridolfi} \\
INFN, Sezione di Genova, Via Dodecaneso 33, I-16146 Genova, Italy\\
\vspace*{0.5cm}
{\bf Abstract}
\end{center}
\noindent
We review the analysis of polarized structure function data using
perturbative QCD at NLO. We use the most recent experimental data to
obtain updated results for polarized parton distributions, first
moments and the strong coupling. We also discuss several theoretical
issues involved in this analysis and in the interpretation of its results.
Finally, we compare our results with other similar analyses
in the recent literature.\\
\vspace*{0.5cm}
\begin{center}
{\it Talks given by G.~Altarelli and G.~Ridolfi at the\\
Cracow Epiphany Conference on Spin Effects in Particle Physics,\\
January 9--11, 1998, Cracow, Poland.}
\end{center}

\vfill
\noindent

\begin{flushleft} CERN-TH/98-61  \\ March 1998 \end{flushleft}
\vfill\eject

\setcounter{page}{1} \pagestyle{plain}

\section{Introduction}

The interest in polarized deep inelastic scattering was revived in
1988 by the results of the famous EMC experiment~\cite{EMC} that led
to the so-called "spin crisis" problem. Since then a lot of progress
has been made. Several experiments were completed at
CERN~\cite{SMCpnew,SMCdnew} and SLAC~\cite{E143new}-\cite{E154nnew} on
proton, deuterium and $^3$He targets. The HERMES experiment is under
way at DESY~\cite{HERMESn}. Other experiments are planned with the
goal of improving our knowledge of spin structure functions: COMPASS
at CERN, HERA with polarized beams at DESY and RHIC at Brookhaven. On
the theory side, the "spin crisis" was immediately recognised not to
be a fundamental problem but rather an interesting property of
spin structure functions to be understood in terms of QCD. A number of
dynamical mechanisms have been proposed and
studied~\cite{altarelliross,AL,efremov,carlitz,instantons,BEK}.  It is
only at the naive parton level that the first moment of the singlet
part of the structure function $g_1$ corresponds to the total helicity
fraction carried in the target by parton quarks. However in
perturbative QCD this identification is no longer valid, even
approximately, due to the effect of the axial anomaly. Extensive
calculations of hard cross sections for polarized processes have been
performed, and in particular the complete two loop evolution kernels
are now available~\cite{NLO}. The Wilson coefficients for the singlet
and nonsinglet first moments of $g_1$ are known up to three
loops~\cite{NNNLO}. By now the perturbative $Q^2$ dependent effects
are computable at the same level of accuracy for polarized and
unpolarized structure functions. In addition, the study of the
expected behaviour of polarized structure functions at small $x$ has
much progressed, in parallel with similar results for unpolarized
parton densities motivated by the HERA experimental data.

At present, we can say that one phase of the study of polarized
structure functions has been concluded. In this first phase attention
was mainly concentrated on first moments and sum rules. The main
conclusions are that the Bjorken sum rule is valid within one standard
deviation while the Ellis-Jaffe sum rule is violated at a level of
about three standard deviations. Attention
is now shifting to the reconstruction of
polarized parton densities at all $x$ and $Q^2$. In particular one is
interested in the gluon density, since this is expected to have
special properties in polarized deep inelastic scattering. The
possibility of inferring the gluon density from scaling violations
is under active study. Of special interest is the behaviour at small
values of $x$ of all polarized parton densities and their variation
with $Q^2$. The data gathered at HERA on the small $x$ behaviour of
unpolarized structure functions and related theoretical work
together imply that a simple Regge extrapolation at small $x$ is
unreliable at large $Q^2$. This Regge extrapolation was used in the
past to derive first moments of polarized structure functions and the
values obtained are indeed significantly biased by this
assumption. The lesson from the HERA experimental results is that
Regge behaviour in general only applies at small
$Q^2\lsim\Lambda^2$. At large $Q^2$ the behaviour induced by the QCD
evolution prevails if it is more singular than the Regge
prediction. Actually, for $x\rightarrow 0$, the Regge behaviour is
less singular than the QCD evolution for all parton densities except
the nonsinglet unpolarized quark densities. The prejudice that Regge
behaviour should be valid at small $x$ for all $Q^2$ values is often
supported by the observation that it has some empirical success in the
case of nonsinglet unpolarized quark densities. But, as we said, this
case is the exception rather than the rule.

\begin{figure}
\begin{center}
    \mbox{
      \epsfig{file=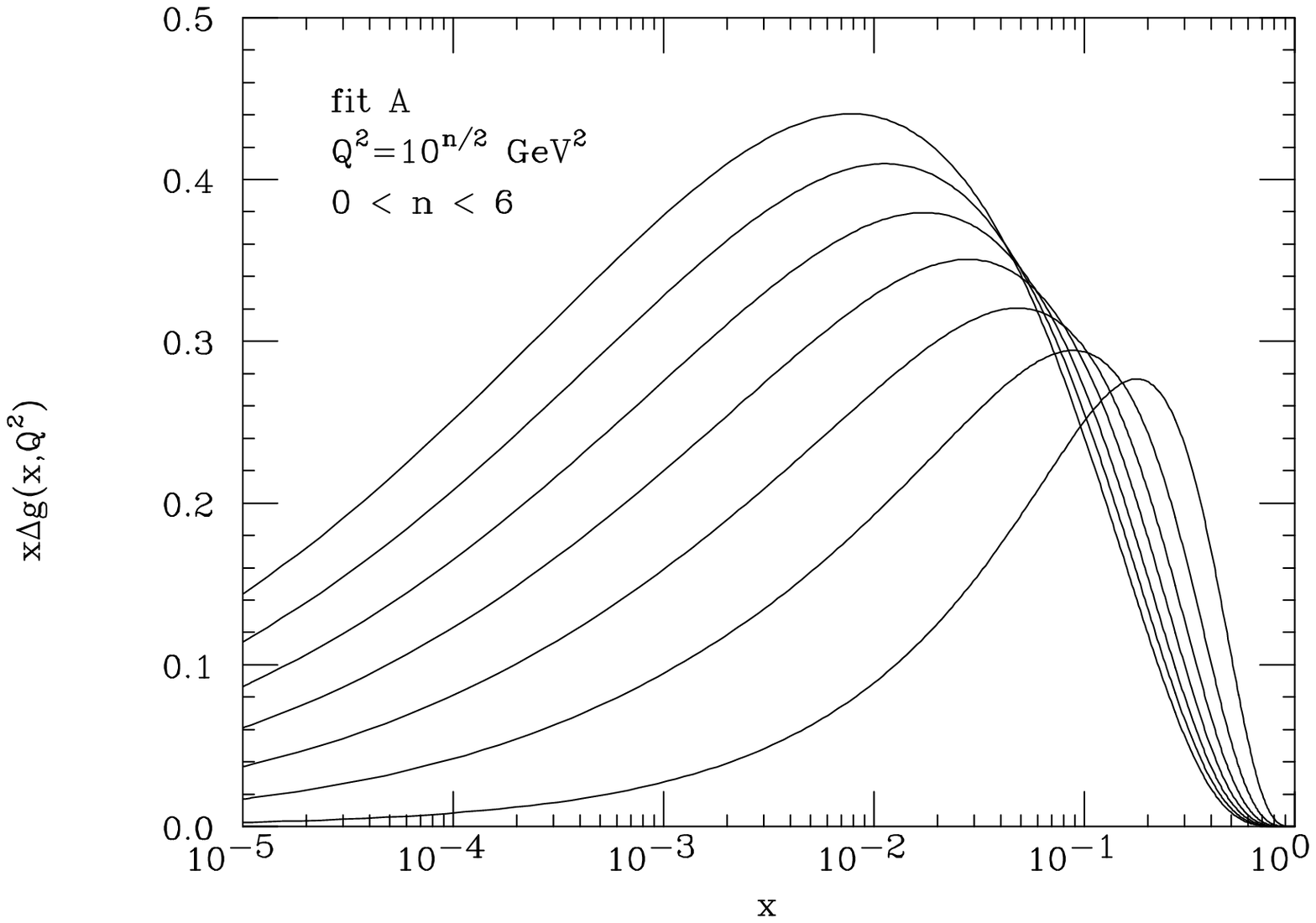,width=0.70\textwidth}
      }
\ccaption{}{
Plot of $x \Delta g(x,Q^2)$ for fit A (see sect.~3). The curves correspond to
$Q^2=10^{n/2}$~GeV$^2$, $n=0,\ldots,6$
}
\label{gluon}
\end{center}
\end{figure}

The gluon density is of special interest in polarized deep inelastic
scattering because its first moment is predicted to increase like
$1/\alpha_s(Q^2)$ while higher moments are decreasing functions of
$\log{Q^2}$ (falling at a faster rate than for the unpolarized gluon density).
The resulting behaviour in $x$ and $Q^2$ is shown in fig.~\ref{gluon}.
The distribution, with growing area, is rapidly shifted to smaller $x$
as $Q^2$ increases. The $Q^2$ behaviour of the first moment of the
polarized gluon density was originally derived when the QCD evolution
equations were first written down in $x$ space. In fact the first
moment of the polarized gluon splitting function is finite and
proportional to the first coefficient of the beta function, which
establishes the quoted relation with the running coupling
$\alpha_s(Q^2)$. This relation between the $Q^2$ evolution of the
first moment of the polarized gluon density and the running coupling
is induced by the axial anomaly and corresponds to the 
non--renormalisation of the anomalous vertex $\alpha_s F_{\mu\nu}\tilde
F^{\mu\nu}$. 

It is well known that there is no operator that corresponds to the
first moment of the polarized gluon density in the operator light cone
expansion. But the gluon density and its first moment can be precisely
defined in the more general context of the QCD improved parton model
which is the established approach for hard processes. For example, we
can use as a defining measurement the (sub)process
$g~+~proton\rightarrow Higgs~+~X$~\cite{AFR}. In leading order, only
the gluon density in the proton contributes to the cross section:
\beq
\sigma(S,M_H^2)=\int{dx~g(x,M_H^2)\tilde{\sigma}(xS)}
=\frac{C}{S}\,g(\frac{M_H^2}{S},M_H^2)+\dots 
\label{1i}
\eeq 
Here we used the fact that in leading order the partonic cross section
$\tilde{\sigma}$ for $g+g\rightarrow H$ is proportional to a delta
function: $\tilde{\sigma}= C\delta(xS-M_H^2)$, with C a constant. Thus
by adjusting $M_H^2/S$ all values of $x$ can be reached and the first
moment can also be computed. Moreover, since the Higgs is a scalar,
for a polarized gluon on a polarized proton only gluons in the proton
with the same helicity as the external gluon can contribute in any
reference frame where the proton and the incident gluon have collinear
spatial momenta. We can thus separately define $g_+$ and $g_-$ in
terms of cross sections for physical processes, which are necessarily
positive definite. The positivity condition $|\Delta g|\leq g$ is thus
automatically guaranteed by the positivity of the cross
section. However this identification is only valid at LO and, in a
generic factorization scheme, it will be violated at NLO and
beyond. As a consequence, when performing NLO fits, it is wrong to
impose the positivity condition within a generic gluon
definition. This unnecessarily restrictive assumption can lead to very
misleading conclusions, particularly if the starting scale is chosen
small, as is often the case.

We have seen that the polarized gluon density, including its
first moment, can be defined from hard processes outside totally
inclusive deep inelastic scattering. At leading order this definition
is totally unambiguous (provided that the defining hard process is
within the set of observables that obey the factorisation theorem).
The first moment of the polarized gluon density, $\Delta g(1,Q^2)$,
obeys the polarized
evolution equations and increases like $1/\alpha_s(Q^2)$. As a
consequence the definition of the singlet quark first moment becomes
totally ambiguous, because two generic definitions differ by terms of
order $\alpha_s(Q^2) \Delta g$. For the first moment what is formally
a next to leading order correction is potentially of the same size
as the leading term. As a consequence the singlet quark first moment,
defined directly from the structure function $g_1$ (i.e. the one used by the
experiments when the "spin crisis"~\cite{LeaderAnselmino}
was announced) does not have to
coincide with the constituent quark value (the total fraction of the
proton spin carried by quarks). Only for exactly conserved quantities
do the corresponding values for constituent and parton quarks have to
coincide. The first moments of the quark densities are in general only
conserved at leading order by the QCD evolution. But, due to the axial anomaly,
the singlet quark first moment defined from $g_1$ is not conserved in
higher orders. However it can be shown that it is possible to select a
definition of the singlet quark first moment in such a way that it is
conserved at all orders. The relation between the singlet quark first
moments in $g_1$, which we will call $a_0(Q^2)$,
and the conserved definition $\Delta \Sigma(1)$ is given by
\beq
a_0(Q^2)~=~\Delta \Sigma(1)~-~n_f\frac{\alpha_s(Q^2)}{2\pi}
\Delta g(1,Q^2).
\label{2i} 
\eeq
One main challenge for experiments on polarized deep inelastic
scattering is to measure the polarized gluon density and its first
moment with sufficient accuracy to be able to verify the above
relation, that is to check that $\Delta\Sigma(1)$ is indeed compatible
with its constituent value. Note that this constituent value appears
to be significantly less than unity, i.e.  $\Delta\Sigma(1)\sim
0.6$~\cite{JaffeManohar}.

The purpose of this paper is to update our recent analysis~\cite{ABFR}
of all the available data on polarized structure functions in order to
extract the polarized parton densities and their moments for
comparison with theoretical expectations. The differences with respect
to the data sets used in ref.~\cite{ABFR} are the following: we
include the most recent SMC data with proton target~\cite{SMCpnew},
which differ considerably from their older analyses in the small $x$
region; we use here the published SMC deuteron data~\cite{SMCdnew} and
E154 neutron data~\cite{E154nnew}, which are slightly different from
the preliminary data sets used in ref.~\cite{ABFR}; and finally we
include the recent data for $g_1$ with a $^3$He target obtained by the
HERMES Collaboration~\cite{HERMESn}. We have not included in our fits
the old data of refs.\cite{EMC,oldSLAC}, which have much larger
uncertainties than those of more recent experiments.  We have checked,
however, that the inclusion of these old data does not affect our
results. Similar NLO analyses have been recently performed by other
groups. We shall comment in the following on these papers, pointing
out the differences with respect to our work.

We will discuss the possibility of obtaining the polarized gluon
density (as well as the other parton densities and $\alpha_s(Q^2)$)
from the observed scaling violations in the $g_1$ data. The result of
our analysis is that there is some indication in the data for a large
positive gluon component, large enough to make $\Delta \Sigma(1)$
close to the constituent value. But the uncertainties are still very
large. In fact our conclusion that the data indicate a large gluon
component is not shared by other authors, as we will discuss in the
following. Let us mention in passing that often the following (false)
argument is made to imply that it is impossible to derive $\Delta g$
or $\alpha_s(Q^2)$ from the present data on scaling violations. The
quantity which is directly measured is the cross section asymmetry
$A_1$ that experimentally does not show any appreciable $Q^2$
dependence within the present accuracy of the data. So -- the argument
goes -- how can it be possible to derive values of $\Delta g$ or
$\alpha_s(Q^2)$ different from zero from data that do not show any
scaling violations? The reason why this argument is false is that
asymptotically $A_1\sim g_1/F_1$, where $F_1$ is the unpolarized
structure function. Now the QCD evolution equations do not apply
directly to $A_1$; rather, as is well known, two different evolution
equations, with different kernels, are valid for $g_1$ and $F_1$. The
approximate cancellation of the scaling violations for $A_1$ in the
measured range is a strong constraint on the $g_1$ scaling violations,
given that the scaling violations for $F_1$ are at present known with
much larger accuracy. Thus it is a remarkable consistency check that
the observed approximate equality of the scaling violations for $g_1$
and $F_1$, when analysed in terms of the evolution equations for
$g_1$, do indeed lead to a value of $\alpha_s(Q^2)$ which is in
agreement with the world average (and many standard deviations away
from zero).

We will also show that when all the data are included it is possible
to make a reliable test of the Bjorken sum rule \cite{Bj}. For an
experimental verification of the Bjorken sum rule one has to extract
from the data the first moment of the difference of polarized up and
down quark densities at some convenient value of $Q^2$. Data taken at
all kinematically accessible values of $x$ and $Q^2$, and on all
available targets, contain information relevant for the reconstruction
of polarized parton densities at a given $Q^2$ and ought therefore to
be included.  The complete NLO evolution kernels \cite{NLO} can be
used to reduce to the same $Q^2$ data measured at different $Q^2$ for
each $x$.  Since the evolution equations \cite{AP} for partons at a
given $x$ and $Q^2$ depend only on the values of the parton densities
at larger values of $x$ and the same $Q^2$, the necessary correction
can only be performed through a general fit to all the data, which
yields a set of polarized parton densities obeying the correct
evolution equations \cite{BFRa,BFRb}. However in order to perform a
fit one must start with a particular ansatz for the parton densities
at some reference $Q_0^2$. Clearly the results of the fit will depend
to some extent on the starting ansatz one adopts, and this dependence
will induce an error in the computed first moments, and in particular
in the Bjorken sum. Here we will devote special attention to this
issue.

Once the data are reduced to a common $Q^2$ for all $x$ values, an
extrapolation to unmeasured values at small and large $x$ is needed in
order to obtain the first moment. The extrapolation at small $x$ is
especially important~\cite{closeroberts}.\footnote{Note that the
behaviour at small $x$ of the input ansatz for the parton densities at
$Q_0^2$ is not relevant for the evolution correction, which only
depends on $x$ values larger than the smallest measured one. On the
contrary the integration at small $x$ that completes a given moment is
very much dependent on the small $x$ behaviour of the input
distributions, as we shall see.} In most of the existing analyses,
including those in most of the experimental papers, it has been performed by
assuming a simple power behaviour based on Regge theory
\cite{Heimann}. This leads to a rather small contribution to first
moments from the small $x$ region, since the expected extrapolation is
at most flat.

As already mentioned, a naive Regge extrapolation is not justified if
one wants to consider first moments in the perturbative region: small
$x$ contributions to first moments can be relatively large, especially
as $Q^2$ increases. Here we will discuss alternative extrapolation
procedures and the errors associated with them. Our guiding principles
will be the validity of Regge predictions at low $Q^2$ and the buildup
with $Q^2$ of the effect of the QCD evolution. From these starting
points we will estimate the uncertainty in the small $x$
extrapolation, which when combined with the evolution corrections and
the more standard sources of error will allow us to quantify the
extent to which the Bjorken sum rule may be tested using existing
data. In practice we will do this by deriving from the data the value
of $g_A$ and the associated error for an appropriate range of values
of $\alpha_s$.

We will then consider the determination of $\alpha_s$ from the
polarized deep inelastic scattering data. Previous attempts in this
direction \cite{EK} have assumed the validity of the Bjorken sum rule,
and used a value for the Bjorken integral obtained from the first
moments given by the various experimental collaborations, and thus
based on naive Regge extrapolation at small $x$. However, when the
effects of perturbative evolution on the small $x$ extrapolation are
properly taken into account, the evaluations of the first moments must
be revised: their errors then turn out to be considerably increased,
and the determination of $\alpha_s$ from the Bjorken sum rule no
longer works so well. However, we are able to show that a much better
determination of $\alpha_s$ may be obtained if all the available data
and not only the Bjorken integral are used in the analysis: the
comparison of the data at small and large $Q^2$ in the measured range
of $x$ then leads to a reasonably precise measurement of $\alpha_s$.

\section{Polarized Structure Functions and Partons}

We begin by summarising various results on the relation between structure
functions and polarized parton distributions, and their behaviour at
small $x$, which will be important for the following discussion.

\subsection{Defining Polarized Parton Densities}

The structure function $g_1$ is related to the polarized quark and gluon
distributions by \cite{BFRb}
\beq
g_1(x,Q^2)=\smallfrac{\langle e^2 \rangle}{2} [C_{NS}\otimes \Delta q_{NS}
+C_S\otimes \Delta \Sigma + 2n_fC_g \otimes \Delta g],
\label{1}
\eeq
where $\langle e^2 \rangle=n_f^{-1}\sum_{i=1}^n e^2_i$, $\otimes$ denotes
convolution with respect to $x$, and the nonsinglet and singlet quark
distributions are defined as
\beq
\Delta q_{NS}\equiv \sum_{i=1}^{n_f}
(\smallfrac{e^2_i}{\langle e^2 \rangle} - 1)
(\Delta q_i+\Delta \bar q_i),\qquad
\Delta \Sigma\equiv \sum_{i=1}^{n_f}(\Delta q_i+\Delta \bar q_i),
\label{2}
\eeq
where $\Delta q_i$ and $\Delta \bar q_i$ are the quark and antiquark
distributions of flavor $i$ and $\Delta g$ is the polarized gluon
distribution. The evolution equations for the polarized parton
densities are given by
\bea
\frac{d}{dt}\Delta q_{NS}
&=& \frac{\alpha_s(t)}{2\pi} P^{NS}_{qq} \otimes \Delta q_{NS},\nonumber\\
\frac{d}{dt}\ \left (\matrix{\Delta \Sigma \cr \Delta g}\right)
&=&\frac{\alpha_s(t)}{2\pi}
\left (\matrix{P_{qq}^S & 2n_fP_{qg}^S \cr P_{gq}^S
& P_{gg}^S }\right) \otimes \left (\matrix{\Delta \Sigma \cr \Delta g}\right),
\label{3}
\eea
where $t = \log{Q^2/\Lambda^2}$. The coefficient functions $C$ and the
polarized splitting functions $P$ are now known at LO \cite{AP} and
NLO \cite{NLO}. Moments of coefficient functions and parton densities
are defined as $f(N) = \int_0^1{dx x^{N-1} f(x)}$ and denoted by
$C(N,\alpha_s)$, $\Delta q_{NS}(N,Q^2)$, $\Delta \Sigma(N,Q^2)$ and
$\Delta g(N,Q^2)$.

As is well known~\cite{revs}, the definition of the singlet quark
density $\Delta\Sigma(x,Q^2)$ must be carefully specified.
In fact, the scheme dependence of its first moment is proportional to
$\alpha_s(t)\Delta g(1,Q^2)$. This implies that the
ambiguity in $\Delta\Sigma(1,Q^2)$ does not vanish asymptotically,
because, due to the axial anomaly, $\alpha_s(t)\Delta g(1,Q^2)$ is scale
independent at LO. 
For a sensible comparison with the constituent quark spin
one must thus define $\Delta\Sigma(1,Q^2)$ in such a way that it is
scale independent \cite{altarelliross,AL}:
$\Delta\Sigma(1,Q^2)$=$\Delta \Sigma(1)$. This is the definition we
will adopt here. We then have
\beq
\Gamma_1(Q^2)\equiv \int_0^1{dx g_1(x,Q^2)}=\smallfrac{\langle e^2 \rangle}{2}
[C_{NS}(1,\alpha_s(t)) \Delta q_{NS}(1)
+C_S(1,\alpha_s(t)) a_0(Q^2)],\label{gammaone}
\eeq
with $a_0$ the singlet axial charge: \beq
a_0(Q^2) = \Delta \Sigma(1) - n_f \smallfrac{\alpha_s(t)}{2\pi}
\Delta g(1,Q^2)
\label{5}
\eeq
The higher moments of the singlet quark distribution are also scheme
dependent, although in a less dramatic way. Various schemes were
discussed in ref.~\cite{BFRb,FBR} and the dependence of the results of
the analysis on the choice of scheme was studied. Here we do not come
back to this issue but instead simply adopt the AB scheme as defined
in ref.~\cite{BFRb}.

\subsection{Small $x$ Behaviour}
In view of the need to extrapolate the data to $x=0$ in order to
compute moments, it is important to summarise the current
understanding of the small $x$ behaviour of structure functions. For
the unpolarized singlet quark and gluon distributions the QCD
evolution equations (\ref{3}) lead to the following asymptotic
behaviour at small $x$ \cite{DeRuj,BF}:
\bea
x g &\sim \sigma^{-1/2}e^{2\gamma\sigma-\delta\zeta}
\big(1+\sum_{i=1}^n \epsilon^i\rho^{i+1}\alpha_s^i \big),\nonumber  \\
x \Sigma &\sim \rho^{-1}\sigma^{-1/2}e^{2\gamma\sigma-\delta\zeta}
\big(1+\sum_{i=1}^n \epsilon_f^i\rho^{i+1}\alpha_s^i \big),
\label{smallx}
\eea
where $\xi= \log{x_0/x}$,
$\zeta=\log{\left(\alpha_s(Q_0^2)/\alpha_s(Q^2)\right)}$,
$\sigma=\sqrt{\xi\zeta}$, $\rho=\sqrt{\xi/\zeta}$, and the $\epsilon$
terms indicate corrections from the $n$-th perturbative order
(with $n=1$ corresponding to NLO). It follows that the structure functions
$xF_1$ and $F_2$ rise at small $x$ more and more steeply as $Q^2$ increases,
though, for all finite $n$,
never as steeply as a power of $x$. For all other parton
distributions $f$ ($f = q_{NS}$, $\Delta q_{NS},\Delta\Sigma,\Delta g)$ one has
similarly \cite{prise,BFRa}
\beq
f \sim \sigma^{-1/2}e^{2\gamma_f\sigma-\delta_f\zeta}
\big(1+\hbox{$\sum_{i=1}^n$}
\epsilon_f^i\rho^{2i+1}\alpha_s^i \big).  \label{smallxbis}
\eeq
Thus these distributions are less singular by a factor of $x$ than the
singlet unpolarized distributions eq.~(\ref{smallx}), while the higher
order corrections are more important at small $x$ since the exponent
$i+1$ is replaced by $2i+1$; this is because the leading small $N$
contributions to the anomalous dimensions at order $\alpha_s^{i+1}$
are $\left(\alpha_s/(N-1)\right)^i$ in the unpolarized singlet case,
but $N\left(\alpha_s/N^2\right)^i$ for the nonsinglet and polarized
distributions.

The limiting behaviour (\ref{smallx},\ref{smallxbis}) implied by the
evolution equations (\ref{3}) at any finite order in perturbation theory
would change if the series of higher order powers of $\log {1/x}$
were summed to all orders to give a powerlike behaviour in $x$, which
would then overwhelm the leading terms. As is well known~\cite{BFKL},
in the unpolarized singlet channel one may obtain a result as singular
as $x^{-\lambda}$, with $\lambda\sim 1/2$, for $x\Sigma$ and $xg$ by
summation of higher order singularities in the Regge limit of
$x\rightarrow 0$ at fixed $Q^2$ (hence fixed $\alpha_s$). However the
meaning and the value of a fixed $\alpha_s$ are quite ambiguous, and
it is not at all necessary a priori that such a singular behaviour is
of relevance in the measured HERA region. In fact the experimental
results from HERA show no evidence at all for this behaviour
\cite{BF,revHera}. In principle the higher order terms could be more
important for the nonsinglet and polarized distributions due to the
$2i+1$ exponent in eq.~(\ref{smallxbis}) instead of $i+1$ in
eq.~(\ref{smallx}). Indeed summing these `double' logarithmic
singularities \cite{KL,BER} appears to lead to a singular behaviour
$f\sim x^{-\lambda}$ with $\lambda\sim 0.5$ for $q_{NS}$ and $\Delta
q_{NS}$, and $\lambda\gappeq 1$ for the singlet densities (which would
imply that the first moment of the singlet part of $g_1$ is actually
divergent).  If one were to take these theoretical predictions
seriously, the errors in the small $x$ extrapolations considered
below, particularly in the singlet channel, would have to be
considered only as lower bounds. However the summation of `double'
logarithms is even less well founded theoretically than the summation
of `single' logarithms in the unpolarized singlet channel, and we
believe that at present none of these results should be taken
too literally \cite{revHera}.

Another important difference between the small $x$ behaviour of
unpolarized and polarized singlet distributions is that in the
unpolarized case only the gluon anomalous dimension carries the
leading singularity, and consequently the rise in the singlet quark
distribution is driven directly by that of the gluon, while in the
polarized case all the entries in the matrix of singlet anomalous
dimensions are singular, and the polarized singlet quark and gluon
distributions mix. It turns out that the leading eigenvector of small
$x$ evolution is then such that the singlet quark and gluon
distributions have opposite sign, which means in practice that the
singlet component of $g_1$ is driven negative at small $x$ and large
$Q^2$ \cite{BFRa}. Contributions to first moments of $g_1$ from the
small-$x$ tail thus tend to become negative when $Q^2$ is sufficiently
large.

The purely perturbative asymptotic predictions
eqs.~(\ref{smallx},\ref{smallxbis}) only hold when the input
distribution at the starting scale $Q_0^2$ is relatively nonsingular:
if the singularity in the input is stronger than that generated
perturbatively then the input will be essentially preserved by the
perturbative evolution. The rise at small $x$ will then be largely
independent of $Q^2$, rather than becoming steeper as $Q^2$ increases.
If we take the starting scale in the crossover region between
perturbative and nonperturbative dynamics, we can presumably take the
small $x$ behaviour of the input from Regge theory. For unpolarized
distributions the input to the singlet distributions (given by the
pomeron trajectory) is then relatively flat, and indeed the dominance
of the perturbative behaviour (\ref{smallx}) is confirmed by $F_2$
data from HERA \cite{BF,revHera}, while the input to the nonsinglet
(given by the $\rho-\omega$ Reggeon trajectory) is singular, behaving
as $x^{-1/2}$, so it is preserved by the evolution and is consistent
with data from NMC~\cite{NMC} and CCFR~\cite{CCFR}.  For polarized
distributions Regge theory suggests that the form of the input should
be given by the $A_1$ trajectory, and thus flat or even vanishing,
behaving as $x^{0}$--$x^{0.5}$~\cite{Heimann}.

In the following we will consider various scenarios which should cover
the spectrum of reasonable small-$x$ behaviors: either we will assume
a physical picture, inspired by the HERA results, in which we assume
the validity of Regge behaviour at small $x$ in the soft region
(i.e. that at some input scale $Q_0^2\lsim 1$~GeV$^2$ the polarized
densities are flat or vanishing), while at larger $Q^2$ the effect of
NLO perturbative evolution is superimposed (giving a perturbative
growth of the form eq.~(\ref{smallxbis})), or, at the opposite
extreme, we will allow steeper inputs in the nonsinglet sector,
provided they are not more singular than $|\Delta q_{NS}(x,Q^2)|\lsim
x^{-0.5}$ as $x\rightarrow 0$. This picture turns out to be consistent
with the data, and gives a constraint on the allowed growth of
$|\Delta q_{NS}(x,Q^2)|$ in the unmeasured region which in turn limits
the possible ambiguity on the Bjorken sum rule from the small $x$
extrapolation.

\section{Polarized Parton Densities from $g_1$ Data}

We now consider in detail the problem of extracting relevant physical
quantities from the existing data. We devote particular attention to
the study of the dependence of the results on the assumed functional
form of the input parton distributions. For this purpose, we consider
a variety of possible parameterizations of the input, we evolve these
up to the values of $x$ and $Q^2$ where data are available by solution
of the evolution eqs.~(\ref{3}) at NLO, and we determine the free
parameters of the input by a best fit of $g_1(x,Q^2)$ eq.~(\ref{1}) to
all the data of refs.~\cite{SMCpnew}-\cite{HERMESn} with $Q^2\ge
1$~GeV$^2$. Experimental data for $g_1$ are obtained from the
experimentally measured asymmetries $A_1$ using a parameterization of
the measured unpolarized structure functions $F_2$ \cite{NMCF2} and
$R$ \cite{SLACR}, consistently neglecting all higher twist
corrections, and, for deuterium and helium targets, accounting for
effects due to the nuclear wavefunction (but not Fermi motion or
shadowing) by a simple multiplicative correction~\cite{corrnuc,E142}.
Throughout this section we take
$\alpha_s(m_Z)=0.118\pm{0.005}$~\cite{alf} and
$a_8=0.579\pm0.025$~\cite{CRb} where $a_8$ is the SU(3) octet axial
charge (in the proton, below charm threshold, $\Delta
q_{NS}(1)\equiv\eta_{NS}=\frac{3}{4}g_A+\frac{1}{4} a_8$).

To begin with we parameterize the initial parton distributions at
$Q_0^2 = 1$~GeV$^2$ according to the conventional form
\beq
\Delta f(x,Q_0^2) = {\cal N}_f \eta_f x^{\alpha_f} (1-x)^{\beta_f}
(1 + \gamma_fx^{\delta_f})
\label{firstclass}
\eeq
where $\Delta f$ denotes $\Delta q_{NS}$, $\Delta\Sigma$ or $\Delta g$
and ${\cal N}_f$ is a normalisation factor chosen such that the first
moment of $\Delta f$ is equal to $\eta_f$. The signs of all parameters
are left free, including the overall factors $\eta_f$ (although the
data always choose $\eta_f$ to be positive). It is particularly
important to see to what extent the data fix the small $x$ behaviour
i.e. the exponents $\alpha_f$ in eq.~(\ref{firstclass}). The $g_1$
data on neutrons show a strong fall at small $x$, while the deuteron
data are much flatter: thus the fitted nonsinglet quark densities at
small $x$ tend rise while the singlet quarks tend to remain fairly
flat. Starting from a generic input set of densities of the class
eq.~(\ref{firstclass}) we can thus easily end up with $\Delta q_{NS}$
considerably more singular than $\sim x^{-0.5}$. However, the
smallest value of $x$ covered by data is still rather large, so
whether this is actually the case will depend on how the functional
form chosen for the fit extrapolates to very small $x$ the behavior
observed in the last few small $x$ data points.

Indeed, a careful analysis reveals that there is a strong correlation
between $\alpha_f$ and $\delta_f$: one can easily push $\alpha_{NS}$
closer to zero by decreasing $\delta_{NS}$ from unity without
appreciable changes in the quality of the fit. Thus we find that the
existing data do not much constrain the behaviour of the nonsinglet at
asymptotically small values of $x$: even within the simple functional
form eq.~(\ref{firstclass}) one still has a considerable flexibility
in the asymptotic behaviour as $x\rightarrow 0$, and $\alpha_{NS}$ can
be made to vary from values close to zero down to values of order
$\sim -1$ by tuning the parameter $\delta_{NS}$. The resulting
uncertainty coming from the small-$x$ extrapolation is therefore
arbitrarily large, if no assumption is made on the small-$x$ behavior
of the nonsinglet distribution. A reasonable bound to the most
singular admissible behavior, i.e. a lower bound on $\alpha_{NS}$ is
provided by the behavior of the unpolarized nonsinglet distributions,
namely $\sim x^{-0.5}$ for $x\to 0$, which is the hardest Regge
behavior in the nonsinglet channel and it is also the behavior
generated by the summation of double logs discussed in
sect.~2.2. Since we will exhibit below small-$x$ behaviors which are
softer than any power, we consider a fit of this class, denoted as fit~A,
corresponding to such most singular behavior, namely:
\beq
\delta_{\Sigma}=\delta_g=1,
\qquad\delta_{NS}=0.75\;\;{\rm(fixed)},
\qquad\beta_g=4\;\;{\rm(fixed)},
\qquad\gamma_{\Sigma}=\gamma_g\qquad {\rm (fit~A).}
\label{9}
\eeq
The result obtained from this fit are listed in table 1.

Next, in order to discuss less singular inputs, we completely change the
functional form of the input densities (while keeping the initial
scale at $Q_0^2=1$~GeV$^2$). We thus choose an input in which the
rise at small $x$ is at most logarithmic (fit~B):
\bea
\Delta \Sigma &=& {\cal N}_{\Sigma}\eta_{\Sigma}x^{\alpha_{\Sigma}}
\left(\log 1/x\right)^{\beta_{\Sigma}}
\nonumber\\
\Delta q_{NS} &=& {\cal N}_{NS} \eta_{NS}
\left[\left(\log 1/x\right)^{\alpha_{NS}} +\gamma_{NS}x
\left(\log1/x\right)^{\beta_{NS}}\right] \qquad {\rm (fit~B),}
\\
\Delta g &=& {\cal N}_g \eta_g \left[\left(\log 1/x\right)^{\alpha_g}
+\gamma_g x\left(\log 1/x\right)^{\beta_g}\right]\nonumber \label{fitb}
\eea
The small $x$ behavior of this fit is weaker than any power, and thus
in particular compatible with the Regge prediction. Note also that
$\log{1/x}\sim(1-x)$ near $x=1$, so the behaviour as $x\rightarrow 1$
is taken care of by the $\gamma$ terms.
The results we obtain from this fit are again reported in table~1. The
quality of fit~B much better than fit~A, as seen from the $\chi^2$ value.

Although the logs are reminiscent of QCD evolution the functional form
of fit B might perhaps appear a little ad hoc. It is thus interesting
to try to generate the logarithms in a more physical way, by
perturbative evolution. In this spirit we consider another set of
trials, where we start the QCD evolution at a very small scale, $Q_0^2
= 0.3$~GeV$^2$, and fit a function of the form eq.~(\ref{firstclass}).
The choice of such a low scale is simply used as a trick to generate
an effective set of distributions at the value of $Q^2$ at which we
begin to fit the data, i.e. $Q^2 = 1$~GeV$^2$ (data with lower $Q^2$
being still discarded), with the logs piled up in a way entirely
consistent with perturbative evolution. In table~1 we report the
results from a fit with
\beq
\gamma_\Sigma=\gamma_g=\gamma_{NS}=0,\qquad
\beta_g=15\;\;{\rm (fixed)}\qquad {\rm (fit~C).}\label{11}
\eeq
In this class of fits, the large-$x$ behaviour of the gluon
distribution can hardly be determined by the fitting procedure;
therefore, we fixed $\beta_g=15$ at $Q_0^2=0.3$~GeV$^2$, because we
checked that this choice approximately corresponds to a $(1-x)^4$
behaviour of $\Delta g$ at large $x$ and $Q^2$ around 1 GeV$^2$.
The quality of the
fit in the measured region is comparable to that of fit A. Comparing
with the results of fit A, we see that by lowering the initial $Q_0^2$
scale all the exponents in the $x^{\alpha}$ terms have become positive
in qualitative agreement with the idea that naive Regge behaviour is
restored at a sufficiently low scale. Indeed a fit of reasonable
quality is also obtained if we fix all exponents $\alpha_f$ at $Q_0^2
= 0.3$~GeV$^2$ to the limiting value admitted by Regge theory, i.e.
one half (fit D):
\beq
\alpha_f = 0.5,\qquad
\gamma_{\Sigma}=0,\qquad\delta_g=1,\qquad\delta_{NS}=1,\qquad{\rm (fit~D).}
\label{12}
\eeq
The results of this fit are also shown in table~1. The $\chi^2$ is now
much worse, but the physical results do not change much, especially in
the nonsinglet sector (for example the central value of $g_A$ is about
the same in fits C and D). One could presumably optimize the choice of
the initial scale $Q_0^2$ to make the agreement with Regge theory even
better. Note that for all fits A-D the $\chi^2$ per degree of freedom
is below 1, but the fit B is neatly preferred~\cite{soper}, while the
fit~D is much disfavoured in terms of the absolute $\chi^2$ value.
\begin{figure}[ptbh]
  \begin{center}
    \mbox{
      \epsfig{file=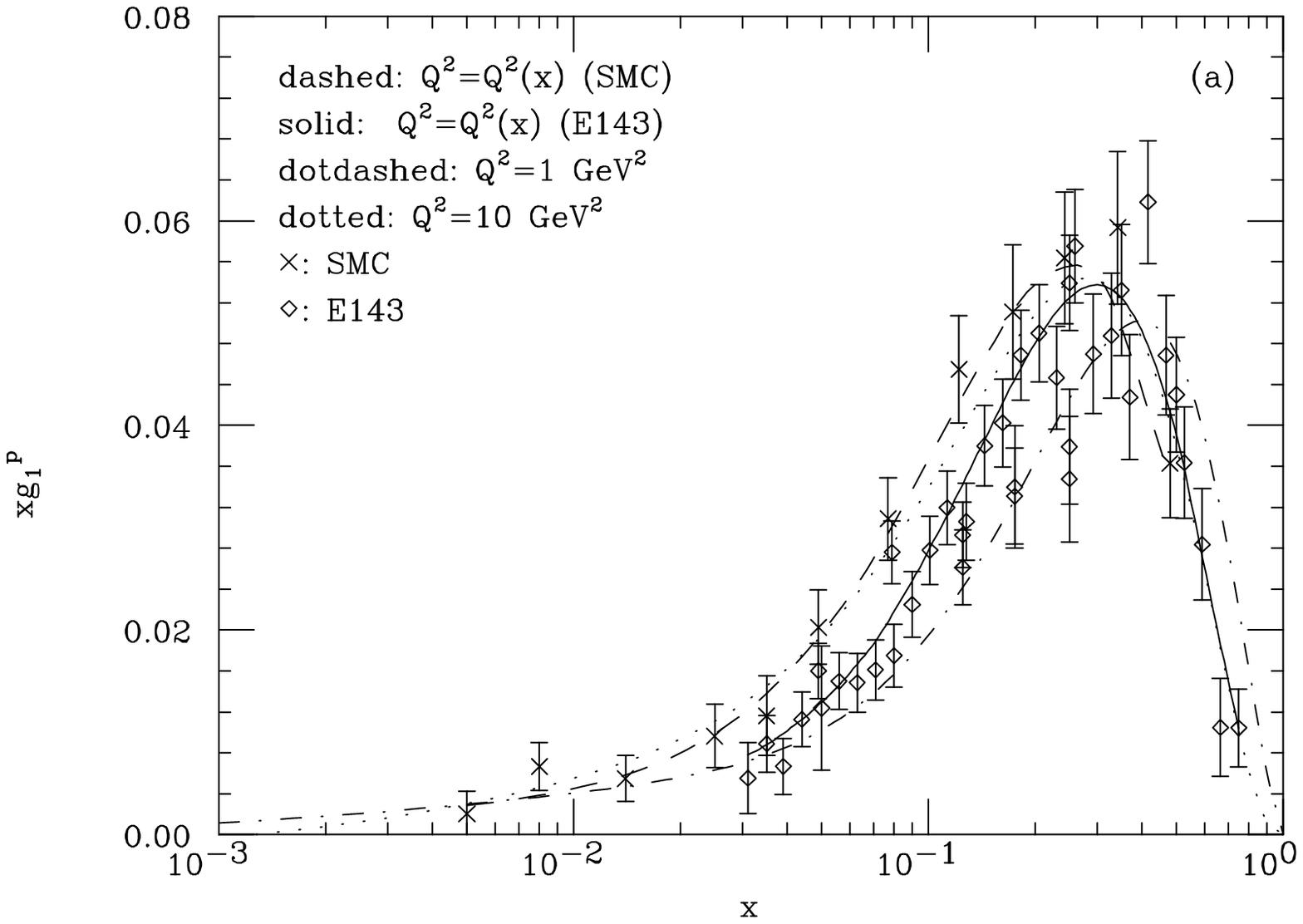,width=0.70\textwidth}
      }
  \end{center}
\end{figure}
\begin{figure}[ptbh]
  \begin{center}
    \mbox{
      \epsfig{file=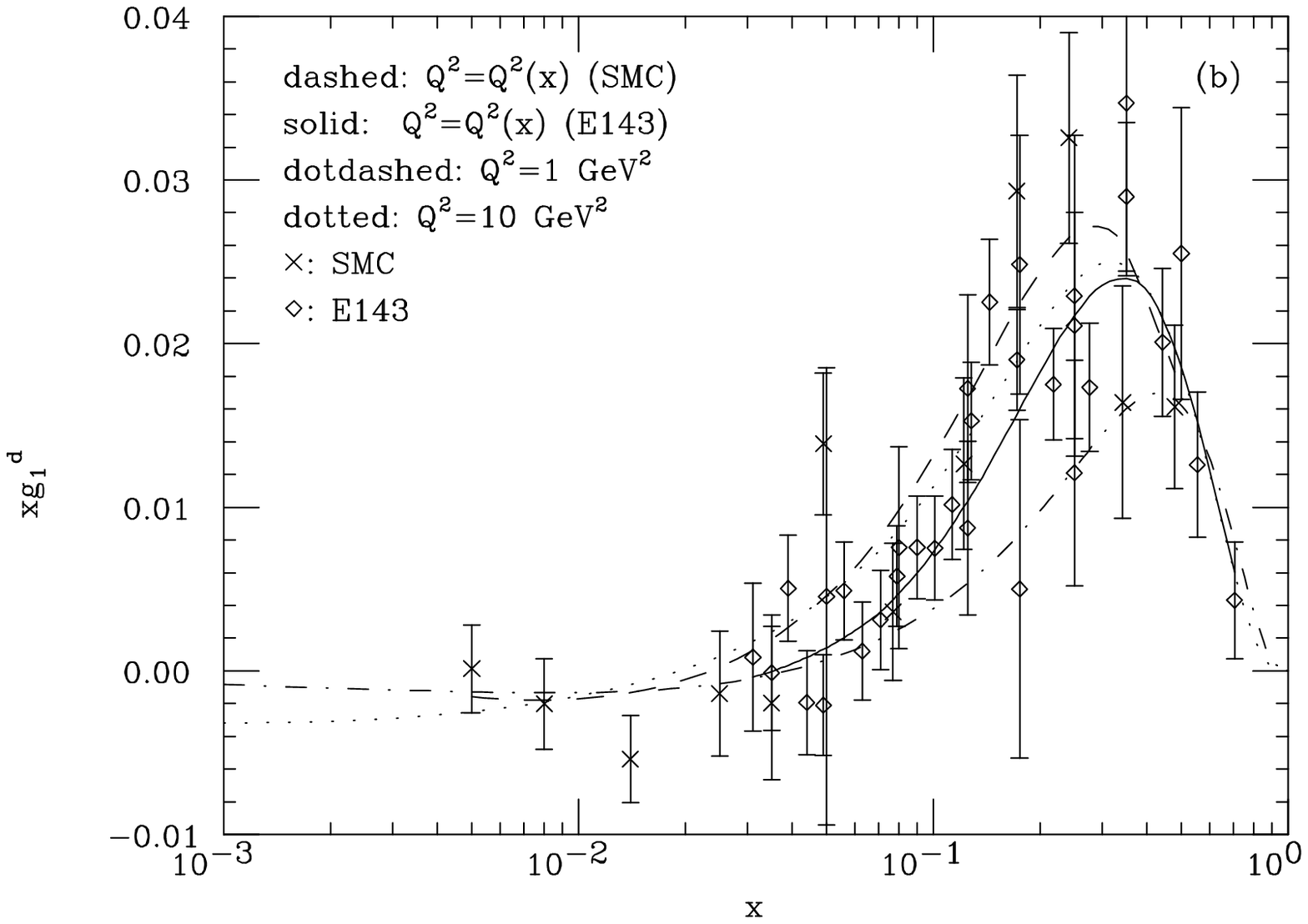,width=0.70\textwidth}
      }
  \end{center}
\end{figure}
\begin{figure}[ptbh]
  \begin{center}
    \mbox{
      \epsfig{file=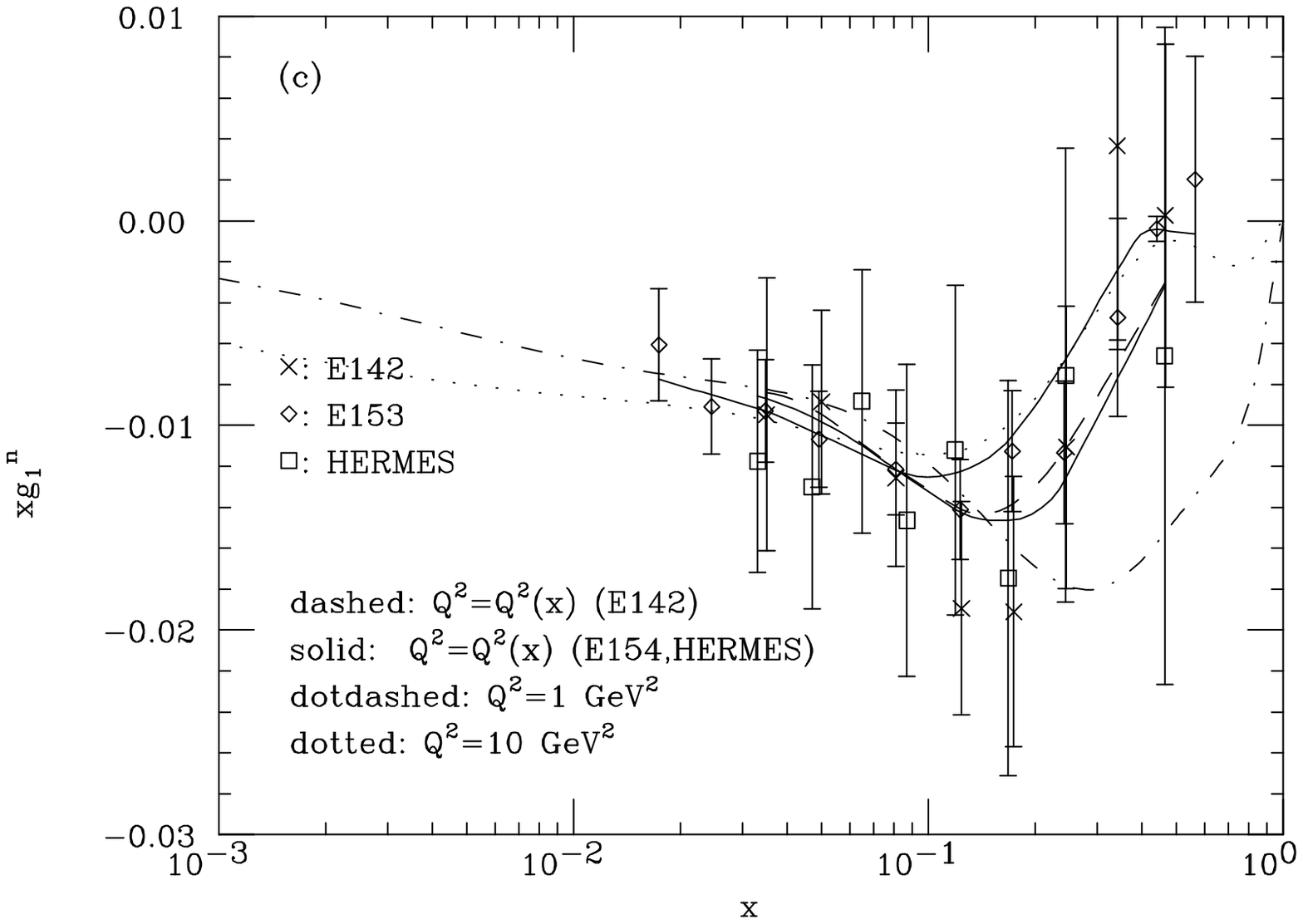,width=0.70\textwidth}
      }
  \ccaption{}{\label{g1plot}
Plots of $x g_1(x,Q^2)$ for fit~B for (a) proton, (b) deuterium, and
(c) neutron targets. The data points with total errors are also shown.
}
  \end{center}
\end{figure}

In figs.~2a-c we display the best-fit $g_1$ (fit B) for protons, neutrons and
deuterons at the $Q^2$ of the data.
\begin{figure}[ptbh]
  \begin{center}
    \mbox{
      \epsfig{file=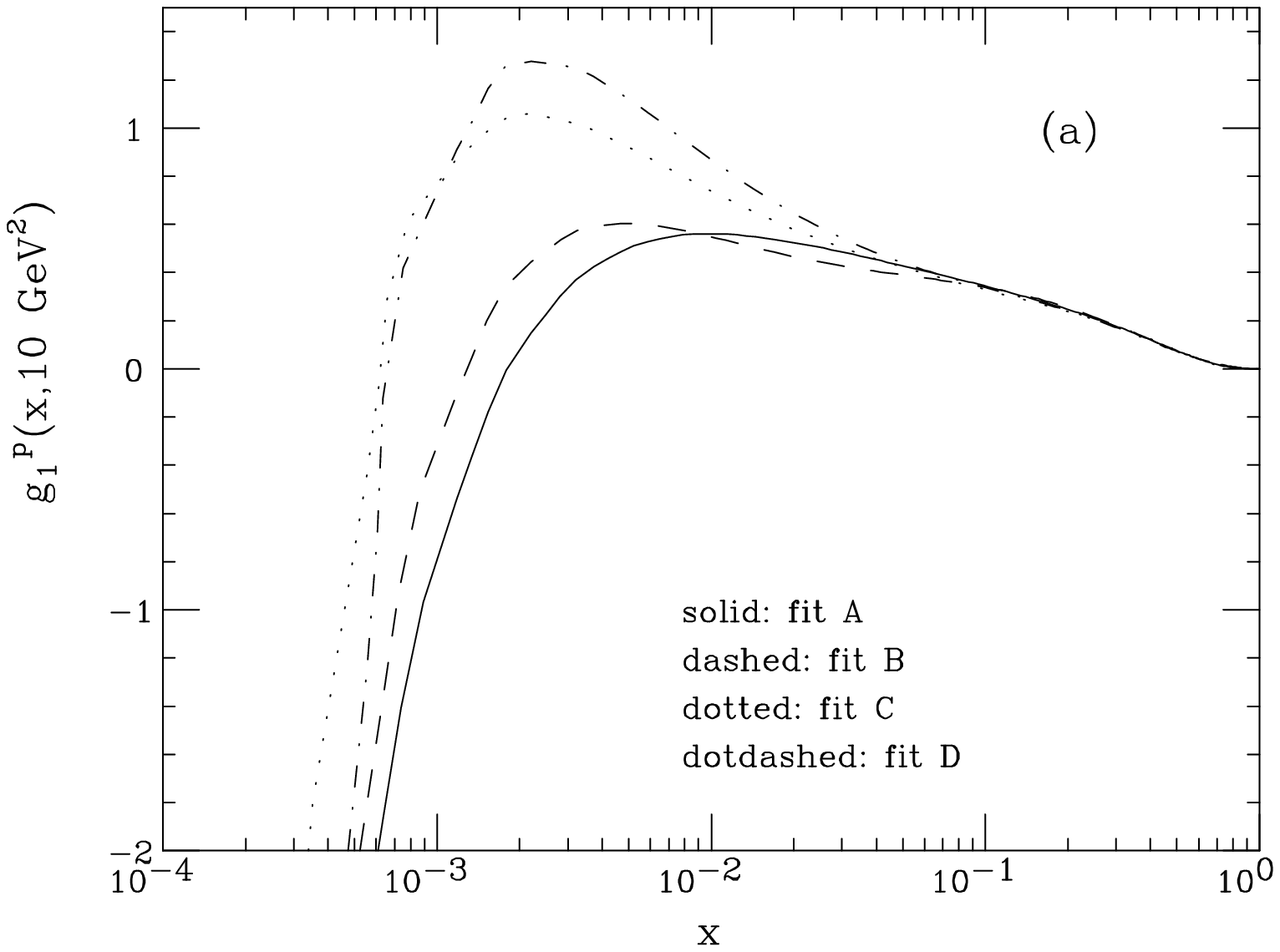,width=0.70\textwidth}
      }
  \end{center}
\end{figure}
\begin{figure}[ptbh]
  \begin{center}
    \mbox{
      \epsfig{file=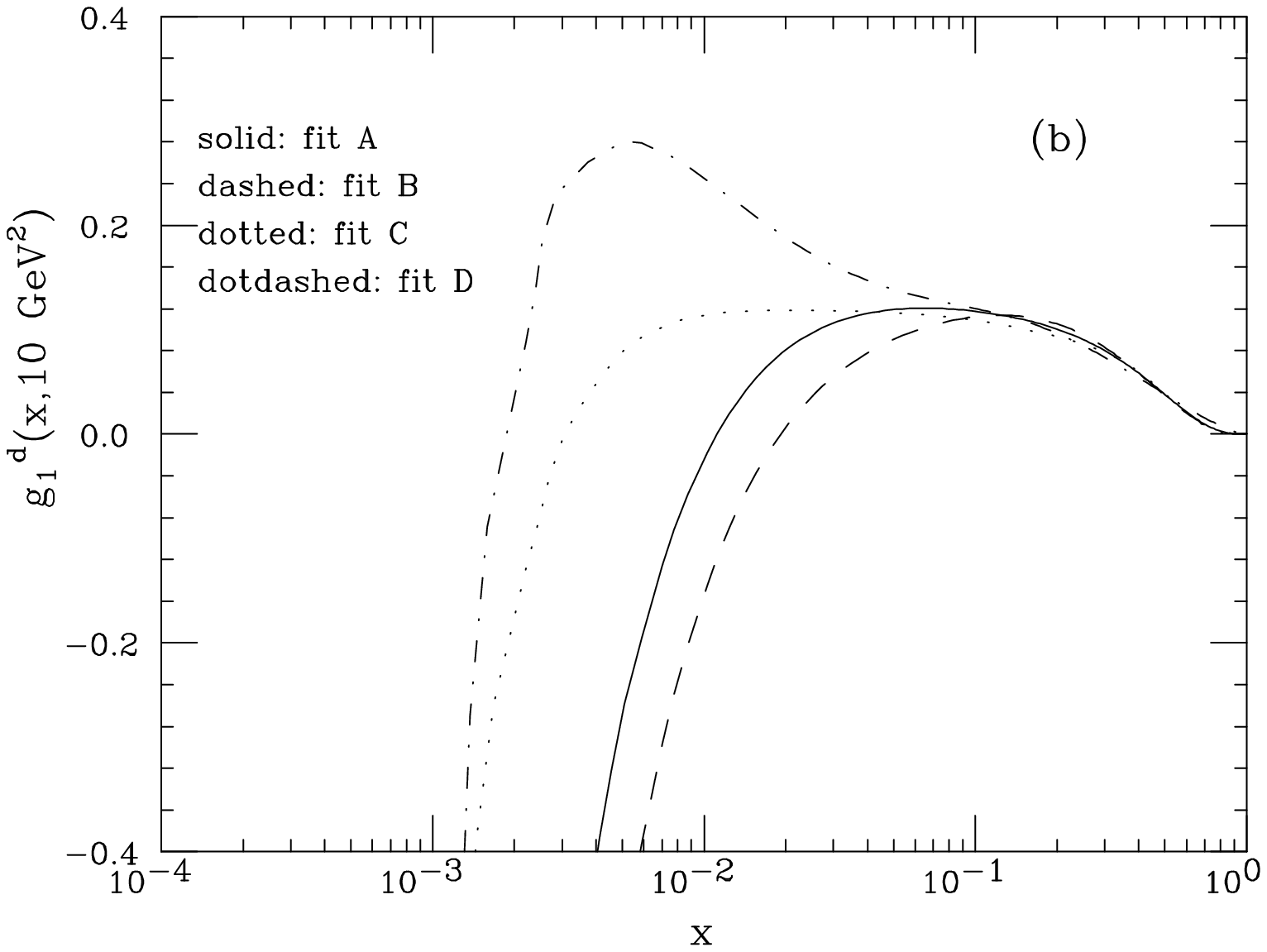,width=0.70\textwidth}
      }
  \end{center}
\end{figure}
\begin{figure}[ptbh]
  \begin{center}
    \mbox{
      \epsfig{file=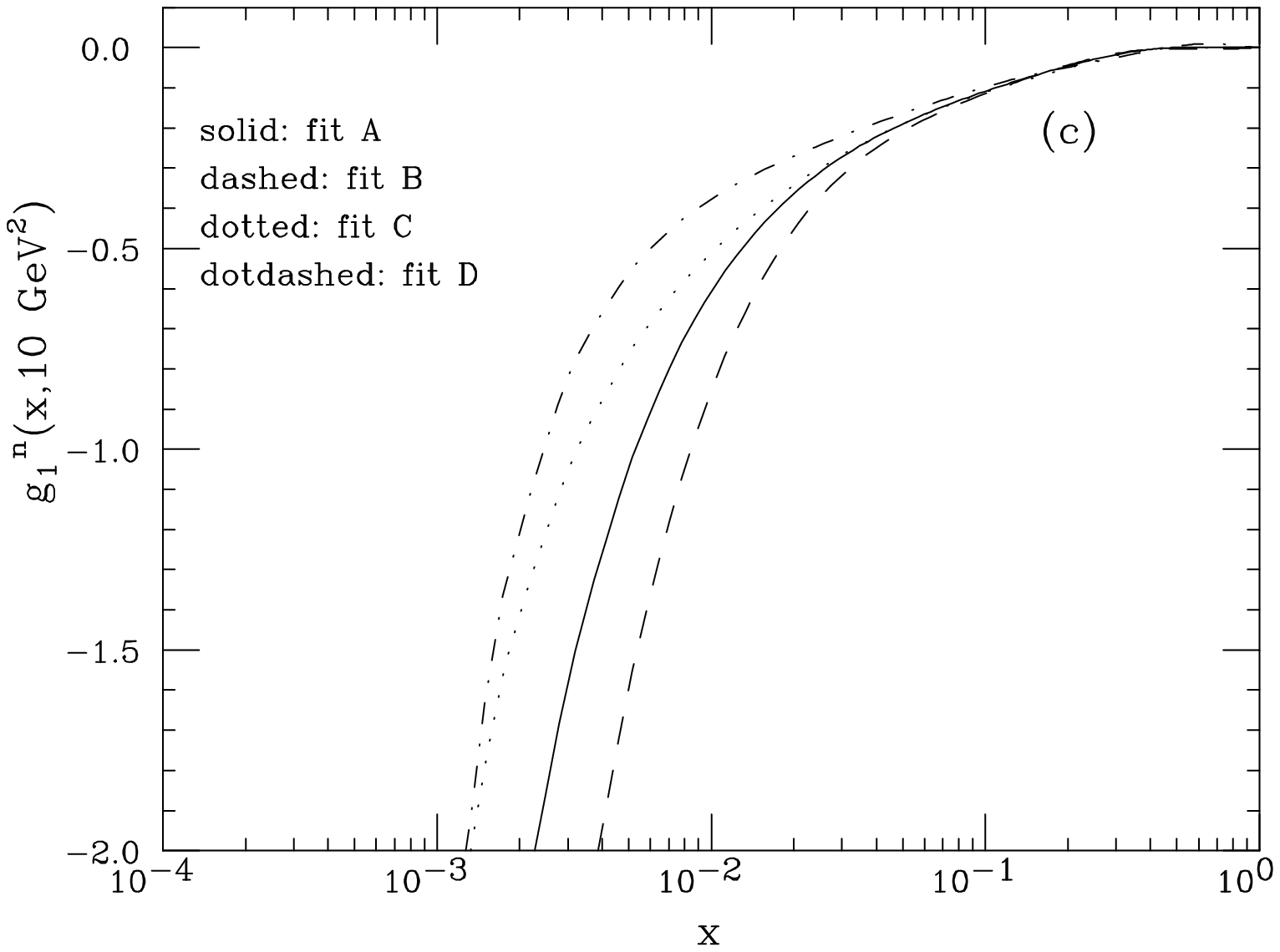,width=0.70\textwidth}
      }
  \ccaption{}{\label{abcdplot}
Plots of $g_1(x,Q^2)$
for (a)  proton (b) deuterium (c) and neutron  targets
 for fits A--D at
$Q^2=10$~GeV$^2$.
}
  \end{center}
\end{figure}
\vfill
\eject
In figs.~3a-c we then compare the best-fit forms of $g_1$
corresponding to fits A--D at $Q^2=10$~GeV$^2$.  The figures show that
while the four fits are reasonably close together in the measured
region ($x\sim0.003-0.03$ up to $x\sim 0.8$), they become very
different in the small $x$ region. Note that at this $Q^2$ value $g_1$
has indeed become negative at small $x$ in all cases. In fig. 4a-c we
display the resulting polarized parton densities obtained from the
fits at the same value of $Q^2$. Here (as in ref.~\cite{ABFR})
$\Delta q_{NS}$ is the quantity defined in eq.~(4) for a deuterium
target, rescaled by a factor $<e^2>=5/18$ (above charm threshold).
Note that the opening of the charm threshold makes the shape of the
nonsinglet different for protons and deuterons, as can be seen
comparing $x\Delta q_{NS}$ (fig.~4b) with
$x\Delta q_3=x(\Delta u+\Delta\bar{u}-\Delta d-\Delta\bar{d})$,
which is displayed in fig.~\ref{uminusd}.
Especially in the singlet sector, the behaviour at small $x$ is
quite different in each case: fits C and D develop a
particularly robust tail at small $x$. It is the
large positively polarized gluon that drives $g_1$ negative at small
$x$.

In table 2 we report the values obtained by computing the first
moments of $g_1$ by integration of the four fits, both in the measured
range of $x$ and in the whole range at the `average' values of $Q^2$
quoted by each experiment on protons, deuterons and neutrons. We see
that while the truncated moments are remarkably close to each other
the complete moments show a much wider spread. We also report the
values of the truncated moments obtained by evolving the data to a
common scale by means of the traditional (but unjustified) assumption
that the asymmetries are scale independent and then summing over the
bins, and those given by the experimental collaborations with their
associated total errors. The latter two values should in principle
coincide, and only differ because of details in the way $g_1$ is
determined from the measured asymmetries (such as the use of different
parameterizations of the unpolarized structure function $F_2$, or the
inclusion of some higher twist corrections, as done by some
experimental collaborations).  The effect of the $Q^2$ dependence in
the measured region is sizeable but smaller than the experimental
error. Much larger is the indirect effect of scaling violations on the
extrapolation at small $x$ because of the larger scale dependence at
small $x$.
\begin{figure}[ptbh]
  \begin{center}
    \mbox{
      \epsfig{file=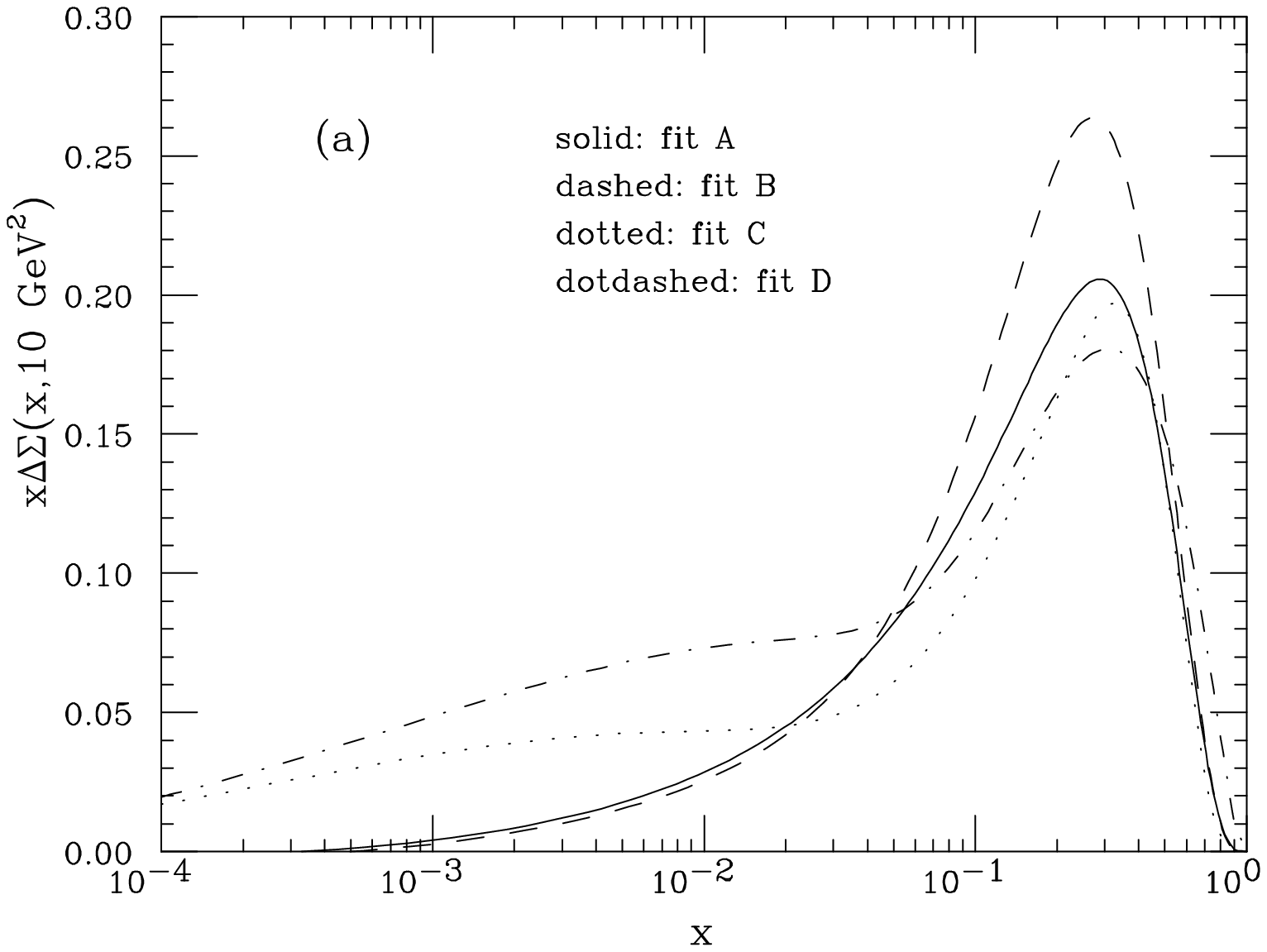,width=0.70\textwidth}
      }
\end{center}
\end{figure}
\begin{figure}[ptbh]
  \begin{center}
    \mbox{
      \epsfig{file=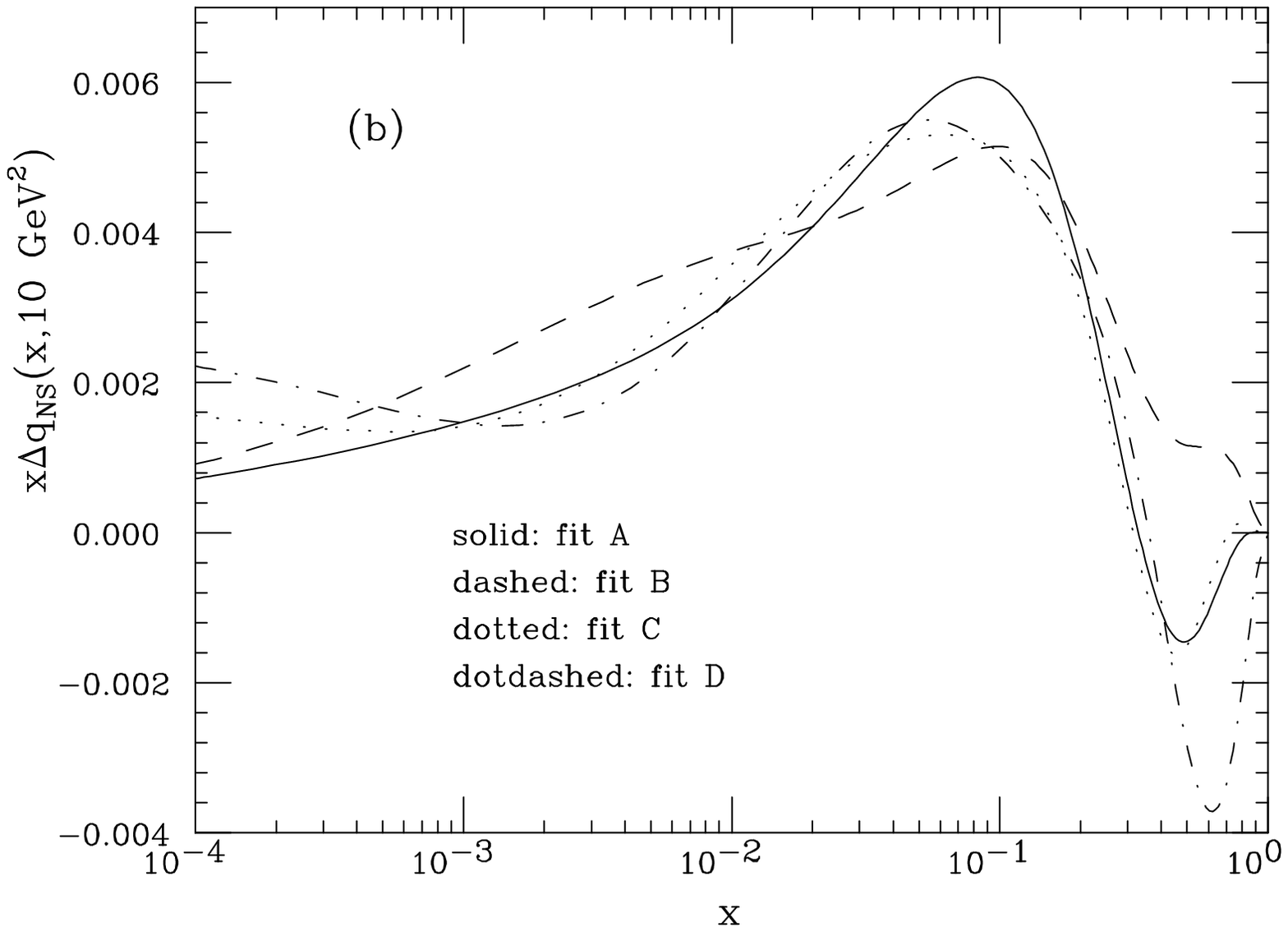,width=0.70\textwidth}
      }
  \end{center}
\end{figure}
\begin{figure}[ptbh]
  \begin{center}
    \mbox{
      \epsfig{file=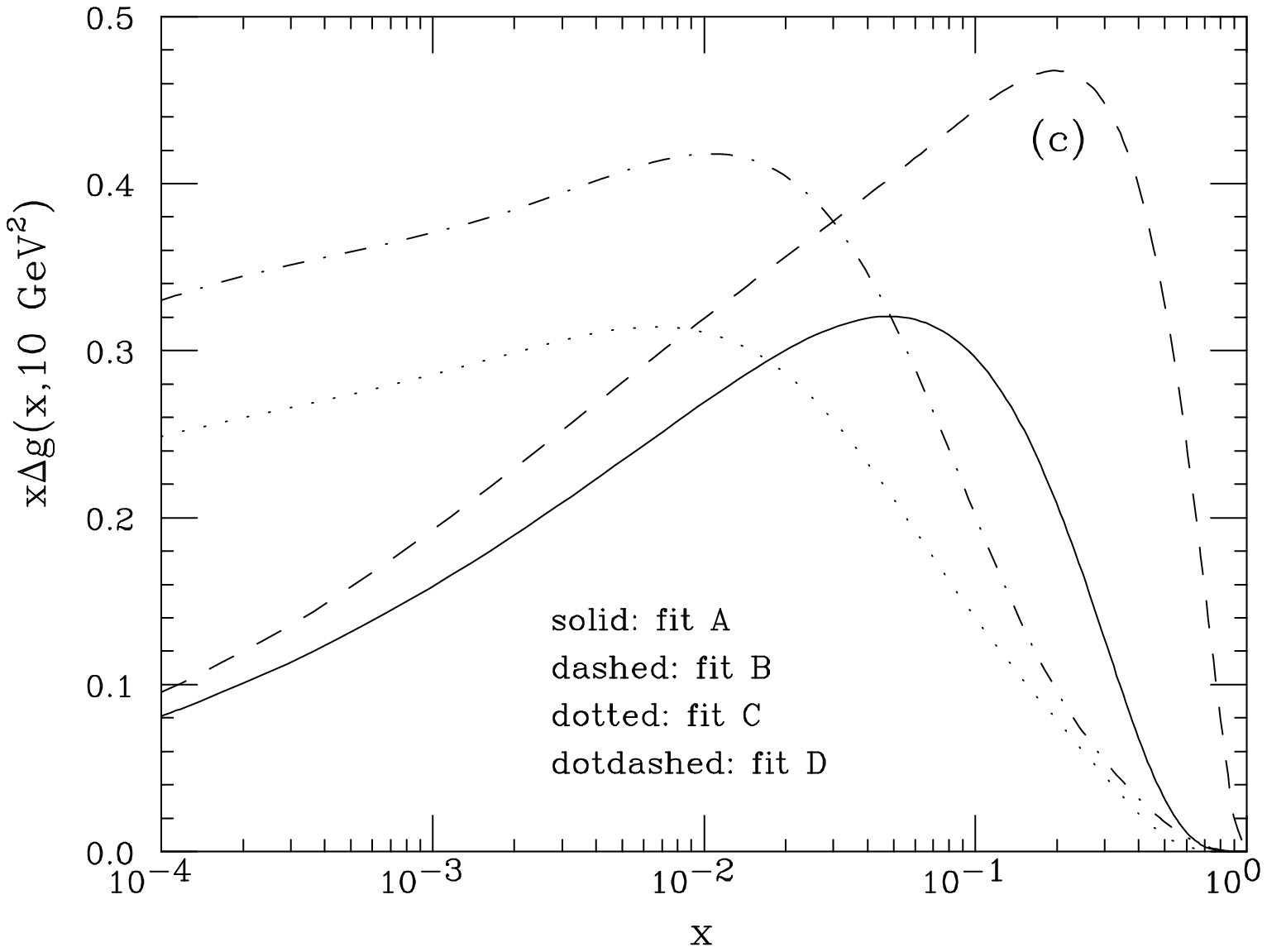,width=0.70\textwidth}
      }
  \ccaption{}{\label{pdfplot}
Polarized quark singlet (a) nonsinglet (b) and gluon (c) distributions
for fits A--D at $Q^2=10$~GeV$^2$.
}
  \end{center}
\end{figure}
\begin{figure}[ptbh]
  \begin{center}
    \mbox{
      \epsfig{file=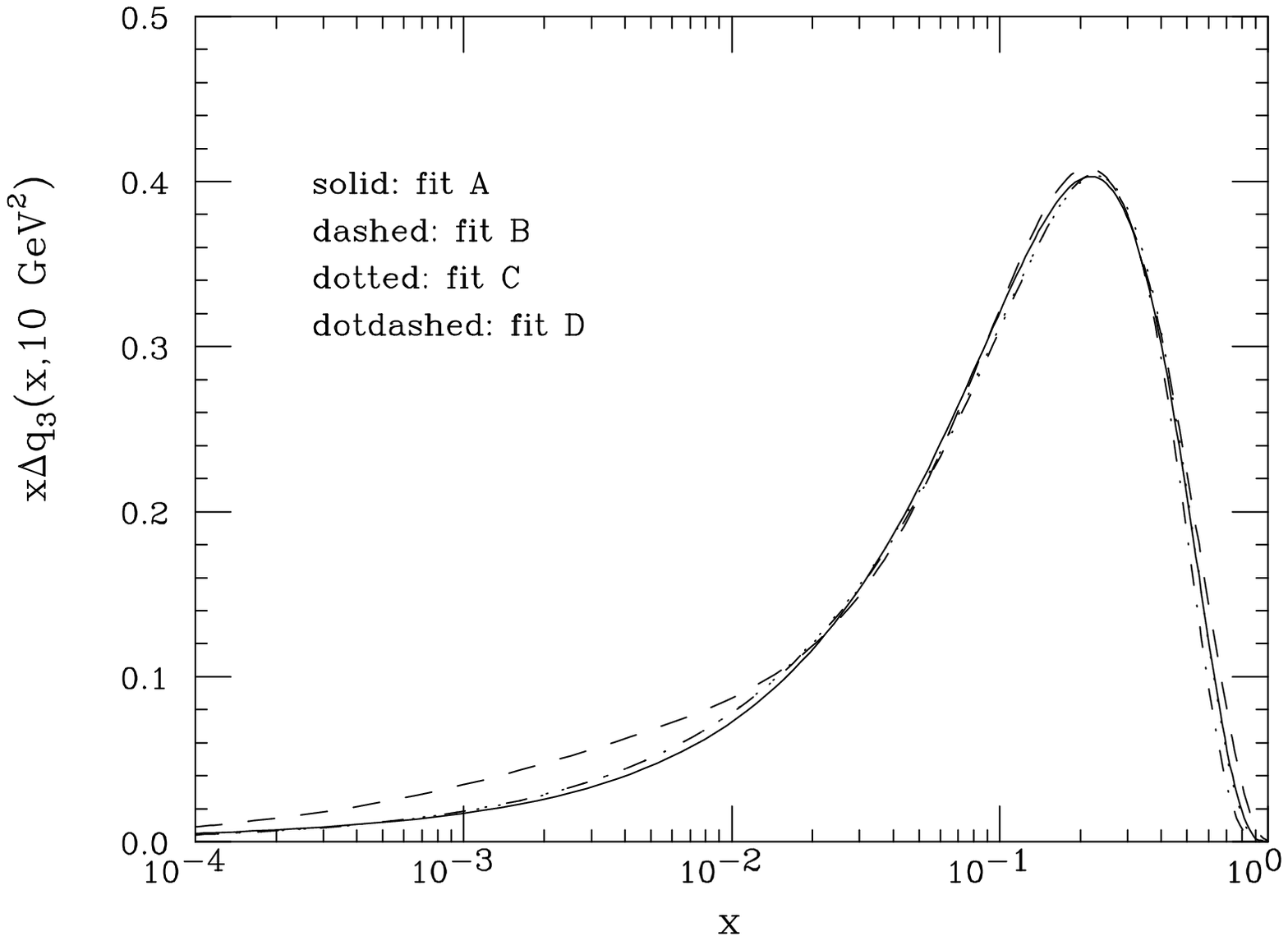,width=0.70\textwidth}
      }
  \ccaption{}{\label{uminusd}
The nonsinglet density
$x\Delta q_3=x(\Delta u+\Delta\bar{u}-\Delta d-\Delta\bar{d})$
in the proton at $Q^2=10 GeV^2$ for fits A-D. The area under the
curves is directly the axial charge $g_A$.}
  \end{center}
\end{figure}

\section{Phenomenological Implications}

We will now discuss the quantitative consequences that can be derived
from the results of the previous section. In particular we will
discuss the status of the Bjorken sum rule, the polarization of
the quark flavor singlet combination and of the gluons in the proton,
the determination of the singlet axial charge, and the
determination of $\alpha_s$. A detailed estimate of the uncertainties
has been given in ref.~\cite{ABFR}. We will not repeat the error
analysis because the quality of the data is essentially unchanged, and
thus use the values of the theory error quoted in ref.~\cite{ABFR}, to
which we refer the reader for details. These values are summarized in
table 3. We wish to point out that we have been particularly careful
in estimating the theoretical uncertainty related to the choice of
functional form adopted for the fits, as well as from the truncation
of the perturbative expansion (renormalization and factorization
scales variation). As shown in table~3, we find that these are the
most important sources of theoretical error.

\subsection{Testing the Bjorken Sum Rule}

One general way to test the Bjorken sum rule is to determine $g_A$ and
the associated error from fitting the whole set of available data
points.  Using the fits described in the previous section and shown in
table 1, and the estimate of the uncertainties of ref.~\cite{ABFR}, we
may take as a result
\beq
g_A=1.18\pm0.05{(\rm exp)}\pm 0.07{(\rm th)}=1.18\pm0.09,
\label{gafit}
\eeq
where the central value is obtained as the average between the maximum
and minimum values of table~1.  This is the procedure we followed in
our previous paper, ref.~\cite{ABFR}. Alternatively, we can take into
account that, contrary to what was found in ref.~\cite{ABFR}, fit B
now has a significantly better $\chi^2$ than any other fit. We can
then take the central value from fit B and introduce an asymmetric
theoretical error to take into account the lower values of $g_A$ from
fits A, C and D. We then obtain
\beq
g_A=1.23\pm0.06{(\rm exp)}\epm{0.06}{0.11}{(\rm th)}
=1.23\epm{0.08}{0.12}.
\label{gafitB}
\eeq
The fitted value is to be compared with the direct measurement
$g_A=1.257\pm 0.003$~\cite{PDG} from $\beta$-decay. Thus we find that
the Bjorken sum rule is confirmed to within one standard deviation but
still with an accuracy of only about 8\%.

\subsection{Singlet First Moments}

Similarly in the singlet sector one obtains from the data values for
$\eta_q=\Delta \Sigma(1)$ (the conserved polarized singlet quark
density), for $\eta_g=\Delta g(1,1\,{\rm GeV}^2)$ (the first moment of
the polarized gluon density evaluated at $Q^2=1$~GeV$^2$), and for
$a_0(10\,{\rm GeV}^2)$ (the non conserved singlet axial charge defined
implicitly from the singlet part of $g_1$ by
eq.~(\ref{gammaone})). This latter quantity approaches a finite limit
at infinite $Q^2$ because the corresponding anomalous dimension starts
at two loops, and within the present accuracy $a_0(10\,{\rm GeV}^2)$
is equivalent to $a_0(\infty)$. The values for these three quantities
as obtained from our representative fits are reported in table~1. We
then studied in detail, following ref.~\cite{BFRb}, the theoretical
errors from the various different sources: results for these are
listed in table~3.  We find (with the same criteria as for $g_A$ in
eq.~(\ref{gafit}))
\bea
\Delta \Sigma (1) &=& 0.46\pm0.04~{\rm (exp)}\pm0.08~{\rm (th)}
= 0.46\pm0.09,
\nonumber\\
\Delta g(1,1\,{\rm GeV}^2)&=&
1.6\pm0.4~{\rm (exp)}\pm0.8~{\rm (th)} = 1.6\pm0.9,\label{firstmom}\\
a_0(\infty)&=&0.10\pm0.05~{\rm (exp)}\epm{0.17}{0.10}~{\rm (th)}
= 0.10\epm{0.17}{0.11}.
\nonumber
\eea
Alternatively, if we proceed as for $g_A$ in eq.~(\ref{gafitB}), we find
\bea
\Delta \Sigma (1)
   &=&0.44\pm0.04~{\rm (exp)}\pm0.08~{\rm (th)}=0.44\pm0.09,
\nonumber\\
\Delta g(1,1\,{\rm GeV}^2)
   &=&1.4\pm0.3~{\rm (exp)}\pm0.8~{\rm (th)}=1.4\pm0.9,
\label{firstmomB}
\\
a_0(\infty)&=&0.11\pm0.05~{\rm (exp)}\epm{0.23}{0.07}~{\rm (th)}
=0.11\epm{0.23}{0.09}.
\nonumber
\eea

The close compatibility of the results (\ref{firstmom}) and
(\ref{firstmomB}) is a reflection of the remarkable stability of the
first moments in the four independent fits A-D.  The parameter
$a_0(\infty)$ measures the degree of `spin crisis': the singlet axial
charge of the nucleon is still compatible with zero as it was at the
beginning of the story \cite{EMC}. Note that with the naive Regge
extrapolation at small $x$ the experimental result for the axial
singlet charge would be significantly larger, with a much smaller
error: for example in ref.~\cite{AlRi} a value
$a_0(\infty)=0.33\pm0.04$ was quoted. There is also some evidence for
a positive gluon polarization in the nucleon (increasing with $Q^2$ as
$1/\alpha_s(Q^2)$). The amount of gluon polarization is large enough
to allow the first moment $\Delta \Sigma (1)$ of the conserved singlet
quark density to get close to $a_8 \sim 0.58$, which in the absence of
all SU(3) and chiral symmetry breaking effects, could be identified
with the constituent spin fraction \cite{EJ}. This can be seen as a
direct confirmation of the physical explanation of the `spin crisis',
advocated in refs.~\cite{altarelliross,efremov,carlitz}, as due to the
anomaly and well described in terms of the QCD parton model (for more
general possibilities, see refs.~\cite{instantons}). 

Whereas the data allow a good determination of the singlet and triplet
quark components, we have verified explicitly that with present data
it is still not possible to separate off the octet quark component and
thus determine the value of $a_8$. In fact the quality of the fit is
essentially unchanged if $a_8$ is varied by 30\% around its central
value. The effects of separating the octet components will be further
discussed in sect.~5.

\subsection{Determination of $\alpha_s$}

The above discussion on $g_A$ makes it clear that the determination of
$\alpha_s$ from the Bjorken integral is adversely affected by the
increased ambiguity from the small $x$ extrapolation that follows from
the demise of the naive Regge behaviour at small $x$. Fortunately we
find that $\alpha_s$ can be determined directly from the available
data without extrapolation in the small $x$ region if the totality of
the data is taken into account and not just the Bjorken integral. The
value of $\alpha_s$ is then determined by the strong scaling
violations needed to accommodate the difference between the data at
small $Q^2$ from the SLAC experiments and those at larger $Q^2$ from
the SMC in the common range of $x$. While $\Delta g(1,1\,{\rm GeV}^2)$
is mainly fixed by the proton data, $\alpha_s$ is determined by the
difference between proton and neutron. 

To show this we repeated the
fits A-D but fixing $g_A$ to its experimental value and instead
fitting $\alpha_s$. In all cases the central value was found to be
close to $\alpha_s(m_Z)=0.120$. Since the various fits differ
considerably in the unmeasured region, this shows that it is the
behaviour in the measured region that matters. In addition, the
Bjorken integral is appreciably different in the different fits A-D
(and the value of $\Delta g(1,1\,{\rm GeV}^2)$ even more so) but this
difference does not affect $\alpha_s$ very much. Furthermore, the
resolution of the discrepancy between the fitted value of $g_A$
eq.~(\ref{gafit}) and its experimental value would require a much
larger increase of $\alpha_s$ if only the Bjorken integral were
relevant for fixing $\alpha_s$. These results show that $\alpha_s$ is
much better constrained by the overall pattern of scaling violations
than by the Bjorken integral alone.
The theoretical uncertainties that affect this determination of
$\alpha_s$ are listed in table~3. The main source of uncertainty
originates from higher order and higher twist contributions.

We thus obtain finally
\beq
\alpha_s(m_Z) = 0.120\epm{0.004}{0.005}{\rm (exp)}
\epm {0.009}{0.006}{\rm (th)}=0.120\epm{0.010}{0.008}.
\label{alphafit}
\eeq
This reasonably good determination of $\alpha_s(m_Z)$ could still be
improved with better data: it is important to notice that without the
very recent neutron data \cite{E154nnew} the experimental error would
be twice as large. We see no reason why it could not be as good as
the determination from unpolarized data if more data were added and
the experimental errors consequently reduced. However the theoretical
error is already the dominant one; it could also be reduced by more
data because it is made particularly large by the small $Q^2$ values
of the neutron measurements.

\section{Comparison with Related Work}
\newcommand\msbar{\overline{\rm MS}}
Recently, after our previous work in ref.~\cite{ABFR}, a number of NLO
analyses of the available data on polarized deep inelastic scattering
have been published \cite{E154th,Stratmann,Leader,Raedel}.  The
results of ref.~\cite{Raedel} are essentially consistent with
ours. Here we briefly comment on the first two papers and their
conclusion. These two papers have many starting points in common and
their conclusions also look similar.

The E154 Collaboration has presented in ref.~\cite{E154th} its own
complete NLO analysis of all available data. What is particularly
interesting is that their final answers to the main questions seem to
be somewhat different from ours, even though there should be no real
contradiction provided all the various uncertainties are invoked in a
conservative form.

Apart from the determination of $\alpha_s(m_Z^2)$, an issue they do
not address, one first main question is the quantitative assessment of
the validity of the Bjorken sum rule. Then there is the determination
of the gluon density with special emphasis on the magnitude of the
first moment and the question of whether the gluon first moment can
explain the apparent difference between the singlet quark moment
defined from $g_1$ and the naive constituent quark expectation.

In fact their result for $g_A$ is somewhat low: for example, in the AB
scheme (which we also use), they find
$g_A=1.07\epm{0.12}{0.06}$. Similarly the value of the gluon first
moment at $Q^2=5$~GeV$^2$ in the AB scheme is found to be $\Delta
g(1,5\,{\rm GeV}^2)=0.4\epm{1.7}{0.9}$, while in the $\msbar$
scheme\footnote{Notice that in this context the definition of the
$\msbar$ scheme is not unique: in fact, a consistent definition of the
$\gamma_5$ matrix in dimensional regularization requires the
introduction of finite counterterms in order to restore the
conservation of nonsinglet axial currents. Here, as is customary in
the literature, by $\msbar$ we mean the factorization scheme adopted
in ref.~\cite{NLO}. An important difference between this scheme and
the AB scheme is that in the AB scheme $\Delta \Sigma(1)$ is scale
independent, whereas in the scheme of ref.~\cite{NLO} it coincides
with $a_0(Q^2)$. The relation between the singlet quark first moments
in the two schemes is thus given by eq.~(\ref{2i}).}  they obtain
$\Delta g(1,5\,{\rm GeV}^2)=1.8\epm{0.7}{1.0}$ (note that $\Delta g$
should be the same in the two schemes at NLO accuracy).  Translated to
$Q^2=1$~GeV$^2$ their gluon result in the $\msbar$ scheme becomes
$\Delta g(1,1\,{\rm GeV}^2)=1.12\epm{0.5}{0.85}$ compared with our
values given in eq.~(\ref{firstmom}).  Thus their results on the gluon
size are quite inconclusive, while we find a moderate indication for a
large positive gluon.

We have emphasized in our work the importance of using a sufficiently
general input parameterization for the determination of the final
results.  The input form used in ref.~\cite{E154th} is different from
any of ours: there, the polarized densities are assumed to be
proportional to the unpolarized ones with a proportionality factor
given by powers of $x$ and $(1-x)$. The initial value of $Q_0^2$ is
very low: $Q_0^2=0.34$~GeV$^2$. Moreover, the two nonsinglet
combinations $\Delta u-\Delta d$ and $\Delta u+\Delta d-2\Delta s$ are
parametrized independently of each other, while in our framework they
are assumed to have the same shape in $x$. In order to check whether
this can be the origin of the differences between our results and
those of ref.~\cite{E154th}, we have repeated our fits with
independent parametrizations for $\Delta u-\Delta d$ and $\Delta
u+\Delta d-2\Delta s$, keeping however the first moment of $\Delta
u+\Delta d-2\Delta s$ fixed to its measured value, $0.579$. This does
not seem to produce any sizeable effect: for example, fit B gives
\bea
\Delta\Sigma (1) &=& 0.45\;\; (0.44\pm 0.05),
\\ 
\Delta g(1,1\,{\rm GeV}^2)&=&1.5\;\;(1.4\pm 0.5) 
\\ 
a_0(\infty)&=&0.10 \;\;(0.11\pm{0.07}) 
\\ g_A&=&1.23\;\;(1.23\pm{0.05}),
\eea
where we have indicated in brackets the values obtained in the
previous section (with $\Delta u-\Delta d$ and $\Delta u+\Delta
d-2\Delta s$ proportional to each other) and the error corresponding
to our estimate of the total error due to the parametrization (from
table 3).  The quality of the fit is unchanged. We therefore conclude
that introducing a separate parametrization of the octet $\Delta u+\Delta
d-2\Delta s$ does not affect our results. In other words,  by
taking the  octet and triplet to be proportional we had introduced no bias,
and thus the errors on physical quantities coming from differences in the input
parametrizations were correctly estimated in ref.~\cite{ABFR}.

Another important point is the choice of the factorization scheme.
Chosing a different scheme induces an uncertainty in the fitted
parameters which is formally of next-to-next-to-leading order. We
estimated the error originated by the truncation of the perturbative
series by changing the value of the renormalization and factorization
scales (which corresponds to a modification of the finite parts of the
counterterms), and we included this large uncertainty in our estimate
of the total error on each quantity.\footnote{Notice that
factorization scale variations were not considered in
ref.~\cite{E154th}, so that the contribution to the theoretical error
due to scheme dependence was not included there.}  In order to check
that this estimate is correct, we repeated our fits in the $\msbar$
scheme, and we found that indeed the central values of the relevant
quantities are within the estimated uncertainties. For example, fit~B
in the $\msbar$ scheme gives
\bea
\Delta g(1,1\,{\rm GeV}^2)&=&1.2\;\;(1.4\pm 0.6)
\label{MSbardg}
\\
a_0(\infty)&=&0.15 \;\; (0.11\epm{0.15}{0.07})
\label{MSbara0}
\\
g_A&=&1.23\;\;(1.23\pm 0.03),
\label{MSbarga}
\eea
where again we have indicated in brackets the values of our original
fit~B, and the estimated error due to scheme change (from table 3). 
Notice that, since in the $\msbar$ scheme
$\Delta\Sigma(1,Q^2)=a_0(Q^2)$, eq.~(\ref{MSbara0}) shows that the first
moment of the quark distribution found in the two schemes agree to NLO
accuracy provided they are transformed using eq.~(\ref{2i}).  Also in
this case, we do not see any significant deviation due to the
difference in the scheme choice, so again we conclude that, while our
estimate ot the scheme-dependence uncertainty was correct, this is not
the source of the discrepancy with ref.~\cite{E154th}.

The two analyses, however, do not only differ in the choice of input
ansatz.  We observe a number of features in the analysis of
ref.~\cite{E154th} that we find unconvincing. First they apply the
positivity relations $|\Delta f|\leq f$ down to their very low initial
value of $Q_0^2=0.34$~GeV$^2$.  We have already remarked that the
positivity relations are not generally true at NLO and the error
induced by using them might in principle be very large at small
$Q^2$. What are bounded by positivity constraints are a number of
physical quantities like asymmetries or cross sections. Only at
leading order do the bounds on these physical quantities translate
into the simple inequality $|\Delta f|/f\leq 1$. At small $Q^2$ there
are large corrections that depend on the precise definition of the
parton densities. The authors of ref.~\cite{E154th} find that the
positivity constraint is very powerful. In our opinion this means that
they have introduced a large bias. Other possible biases are
introduced by the fact that they separate valence and sea, which
cannot, even in principle, be disentangled by measurements of $g_1$;
this separation is, so to say, inherited from the assumed
proportionality to unpolarized densities. Furthermore their
parametrisation implies that, in the interesting region of parameters,
the singlet quarks necessarily dominate over the nonsinglet, while we
find the opposite. Although they get a reasonably good fit in the
measured region, still the fact that we show that a different region
of lower minimum $\chi^2$ exists which they cannot access shows again
that a bias has been introduced. Also, they use a value of
$\alpha_s$ which is very low and by now completely obsolete:
$\alpha_s(m_Z)=0.109\epm{0.007}{0.001}$.  We have checked explicitly
that such a small value of $\alpha_s$ induces an overestimate of the
singlet axial charge $a_0$, and an underestimate of $g_A$. For all
these reasons we are not convinced that the E154 fit is really so
significant or representative.

The work in ref.~\cite{Stratmann} is a summary and update along the
lines of previous analyses~\cite{GRSV}. This analysis has some
features in common with the E154 approach we have just described. In
fact the same parametrization of the polarized densities as
proportional to the unpolarized ones is adopted, and the initial scale
is the same: $Q_0^2=0.34$~GeV$^2$, thus again, in particular,
enforcing a positivity constraint down to this very low initial scale.
Therefore, their fit suffers from the same unjustifiable bias as that
of E154.  Furthermore, the fits of ref.~\cite{Stratmann} have been
performed including also data points with 0.6~GeV$^2\leq
Q^2\leq$1~GeV$^2$. This is undeniably questionable in the context of a
perturbative analysis, especially because data at small $Q^2$ are
usually taken in the small-$x$ region, where the effect of evolution
is very important.

The issues of testing the Bjorken sum rule and of measuring $\alpha_s$
were not addressed in ref.~\cite{Stratmann}. The Bjorken sum rule is
imposed, and a fixed value of $\alpha_s(m_Z)=0.109$ is input
(again very low). The analysis is performed in the $\msbar$ scheme.
What is interesting is that several input forms for the gluon are
studied and compared. For example, one form starts with $\Delta g=g$
at the input scale, another with $\Delta g=0$ and they are compared
with the best fit form of $\Delta g$. With the first option one ends
up with a rather large polarized gluon density while the last one
leads to a negligible amount of polarized gluons. The best fit is
intermediate, with a moderate value of the first moment:
$\Delta g(1,10\,{\rm GeV}^2)\sim 1.45$. The three
$\chi^2$ values are not much different: $\chi^2=127.44, 123.02,
124.24$ for the large gluon, the best fit and the small gluon,
respectively. Thus the small gluon is only about 1~$\sigma$ away from
the best fit. The conclusion of the paper is that the data do not
allow to derive any consequence on the size of the gluon component,
and in particular that a small gluon is perfectly compatible with the
data. 

We observe that their best fit value for the first moment is somewhat
on the low side. However, we believe that several sources of bias have
been introduced in this analysis. We can test this by comparing the
best-fit $\chi^2$ of the present analysis to ours. To this purpose,
notice that the data of refs.~\cite{EMC} and \cite{oldSLAC}, as well
as the low $Q^2$ data which are included in the fits of
ref.~\cite{Stratmann} (40 data points, overall), have very large
errors and thus lower considerably the $\chi^2$ per degree of
freedom. We do not consider the inclusion of low $Q^2$ points to be
acceptable, but if we were to include the data of
ref.~\cite{EMC,oldSLAC} with $Q^2\geq 1$~Gev$^2$ (22 data points) our
best fit (fit B) would drop to $\chi^2/$d.o.f.=0.74, to be compared to
the value $\chi^2/$d.o.f.=0.78 of ref.~\cite{Stratmann}.
Therefore, it is true that, given the size of the existing errors,
a nearly vanishing first moment of $\Delta g$ is within 1~$\sigma$ from
the central value found in ref.~\cite{Stratmann}; however, such a small
central value appears to be considerably biased, and indeed it appears to
correspond to a significantly larger~\cite{soper} value of the total
$\chi^2$. Therefore, we can say that if one samples over a larger set of
input parametrisations the central value is actually larger; the result of
\cite{Stratmann} is compatible with ours, but artificially displaced on
the low side.

In summary, while we agree that the determination of the polarized
gluon density from scaling violations is affected by large
uncertainties, we claim that we have examined a wide range of starting
parametrizations in our analysis and thereby avoided some of the more
questionable assumptions and biases that are often found in the
literature. On the basis of our more systematic work we have presented
some moderate evidence for a large and positive gluon component.
Furthermore, we have shown that the value of the singlet axial charge
is affected by a substantially larger error (while the central value
is smaller) than usually claimed.

\section{Conclusion and Outlook}

In the present paper we have presented an update of our recent
analysis of all existing data on the polarized structure function
$g_1$ and a comparison with other similar analyses. We have addressed
the main questions of the validity of the Bjorken sum rule, of the
determination of the strong coupling and of the measurement of the
polarized gluon density. Overall we find a remarkable consistency of
data and theory: the Bjorken sum rule is valid within the existing
errors, the value of $\alpha_s$ extracted from the data is in good
agreement with the world average for this quantity and we find some
evidence for a large and positive gluon component that could perhaps
be large enough to explain the difference between parton and
constituent quarks. It is true however that the determination of the
gluon density from the observed scaling violations is the most
ambiguous and controversial aspect of the analysis. A substantial
improvement of our knowledge of the polarized gluon density could be
obtained through better and more extended measurements of $g_1$ at
small $x$ as could be made at HERA with polarized proton
beams~\cite{DeRoeck}. In addition, more direct information on the
polarized gluon can be obtained from the study of additional hard
processes beyond totally inclusive deep inelastic scattering, as is
planned at COMPASS~\cite{Nassalski}, at HERA with polarized proton
beams and at RHIC. The preparation of this new phase of the study of
polarized structure functions is now actively in progress. Several
different strategies have been discussed for the determination of the
gluon polarization from
experiment~\cite{Feltesse}-\cite{Contreras}. Most of the early
analyses were limited to leading order, but the corresponding work on
extending these analyses at the next to leading order is now in
progress~\cite{Gehrmann}-\cite{DeFlorianVogelsang}. The prospects for
the future are exciting and the field will remain of great interest in
the years to come.

\section*{Acknowledgement}

G.~Altarelli and G.~Ridolfi are very grateful to Professor Marek Jezabek
and to the Local Organising Committee for their kind invitation and very
warm hospitality in Cracow.

\vfill\eject

\vfill\eject

\begin{center}
\begin{table}
\vspace*{0.5cm}
\begin{center}
\begin{tabular} {|l|l|l|l|l|}
\hline Parameters&A&B&C&D\\
\hline
d.o.f. & $123-10$ & $123-11$ & $123-8$ & $123-8$\\
$Q_0^2/\GeV^2$ & 1 & 1 & 0.3 & 0.3\\
\hline
$\eta_{\Sigma}$ & $0.405\pm0.032$ 
                & $0.440\pm0.037$ 
                & $0.420\pm0.022$ 
                & $0.516\pm0.031$\\
$\alpha_{\Sigma}$ & $0.507\pm0.169$ 
                  & $1.726\pm0.433$ 
                  & $3.156\pm1.028$ 
                  & $0.5$~~~(fixed)\\
$\beta_{\Sigma}$ & $3.232\pm0.575$ 
                 & $2.748\pm0.472$ 
                 & $3.764\pm1.332$ 
                 & $0.816\pm0.194$\\
$\gamma_{\Sigma}$ & $3.006\pm1.862$ 
                  & $0$~~~(fixed) 
                  & $0$~~~(fixed) 
                  & $0$~~~(fixed)\\
$\eta_g$ & $0.949\pm0.185$
         & $1.415\pm0.322$ 
         & $0.506\pm0.094$
         & $0.687\pm0.099$\\
$\alpha_g$ & $-0.486\pm0.281$ 
           & $3.174\pm1.494$ 
           & $0.155\pm0.289$
           & $0.5$~~~(fixed)\\
$\beta_g$ & $4$ (fixed) 
          & $1.032\pm0.535$ 
          & $15$~~~(fixed)
          & $11.58\pm5.69$\\
$\gamma_g$ & $3.006\pm1.862$
           & $47.0\pm103.1$
           & $0$~~~(fixed)
           & $-0.541\pm4.896$\\
$g_A$ & $1.140\pm0.043$
      & $1.232\pm0.057$
      & $1.142\pm0.030$
      & $1.121\pm0.029$\\
$a_8$ & $0.579$~~~(fixed)
      & $0.579$~~~(fixed)
      & $0.579$~~~(fixed)
      & $0.579$~~~(fixed)\\
$\alpha_{NS}$ & $-0.576\pm0.049$
              & $1.662\pm0.144$
              & $0.874\pm0.220$
              & $0.5$~~~(fixed)\\
$\beta_{NS}$ & $2.668\pm0.218$
             & $5.399\pm0.208$
             & $2.325\pm0.421$
             & $3.115\pm0.331$\\
$\gamma_{NS}$ & $34.36\pm17.64$
              & $-0.214\pm0.088$
              & $0$~~~(fixed)
              & $9.869\pm8.739$\\
\hline
$\chi^2$ & 96.4 & 90.5 & 97.6 & 110.5\\
$\chi^2$/d.o.f. & $0.853$ & $0.808$ & $0.849$ & $0.960$\\
\hline
$\Delta g(1,1 \GeV^2)$ & $0.95\pm0.18$
                       & $1.41\pm0.32$
                       & $1.67\pm0.31$
                       & $2.20\pm0.32$\\
$a_0(10 \GeV^2)$ & $0.18\pm0.04$  & $0.11\pm0.05$ & $0.04\pm0.04$ &
$0.02\pm0.03$\\
\hline
\end{tabular}
\end{center}
\caption{Results of fits A--D described in the text}
\end{table}
\end{center}

\begin{center}
\begin{table}
\vspace*{0.5cm}
\begin{center}
\begin{tabular} {|l|l|l|l|l|l|l|}
\hline
 & SMC~:~p &E143~:~p & SMC~:~d & E143~:~d & E142~:~n & E154~:~n\\
\hline
$\langle Q^2\rangle /\GeV^2$ & 10 & 3 & 10 & 3 & 2 & 5 \\
\hline
Meas.~Range:~Exp. & 0.139 & 0.1200 & 0.0407 &
0.0400 & $-0.0280$ & $-0.0360$\\
Exp.~Error &$\pm0.01$ & $\pm0.0089$ & $\pm0.0069$ &
$\pm0.0050$ & $\pm0.0085$ & $\pm0.0064$\\
\hline
$A_1$ ind. $Q^2$ & 0.1478 & 0.1041& 0.0443& 0.0394& $-0.0290$ & $-0.0362$\\
\hline
Meas.~Range:~A & 0.1257 & 0.1080 & 0.0444 & 0.0393 & $-0.0317$ & $-0.0319$\\
Meas.~Range:~B & 0.1247 & 0.1039 & 0.0404 & 0.0346 & $-0.0361$ & $-0.0364$\\
Meas.~Range:~C & 0.1276 & 0.1085 & 0.0458 & 0.0402 & $-0.0293$ & $-0.0318$\\
Meas.~Range:~D & 0.1309 & 0.1079 & 0.0505 & 0.0410 & $-0.0304$ & $-0.0287$\\
\hline
Full~Range:~A & 0.1175 & 0.1142 & 0.0301 & 0.0292 & $-0.0551$ & $-0.0566$\\
Full~Range:~B & 0.1169 & 0.1130 & 0.0224 & 0.0210 & $-0.0703$ & $-0.0716$\\
Full~Range:~C & 0.1021 & 0.0981 & 0.0145 & 0.0128 & $-0.0720$ & $-0.0728$\\
Full~Range:~D & 0.0970 & 0.0926 & 0.0110 & 0.0089 & $-0.0745$ & $-0.0749$\\
\hline
\end{tabular}
\end{center}
\caption{Determination of the first moment $\Gamma_1(\langle Q^2\rangle)$
eq.~(\ref{gammaone}).
For each experiment we display the average value of $Q^2$ and the
contribution to the first moments from the measured range of $x$,
as given, first, by the experimental collaborations, with the
corresponding total (statistical and systematic) error,
then by summing over experimental bins while evolving the data
assuming  scale independent asymmetries, and finally as obtained from
integration of
the fits A--D. In the last four rows the complete first
moments obtained from the fits A--D are shown.}
\end{table}
\end{center}

\begin{center}
\begin{table}
\vspace*{0.5cm}
\begin{center}
\begin{tabular} {|l|l|l|l|l|l|l|}
\hline  & $g_A$ &$\Delta\Sigma$& $\Delta g$
&$ a_0$ & $\alpha_s$\\
\hline
experimental  & \apm0.05 & \apm0.04 & \apm0.4 & \apm0.05 &
\epm{0.004}{0.005} \\
\hline
fitting       & \apm0.05 & \apm0.05 & \apm0.5 & \apm0.07 & \apm0.001 \\
$\alpha_s$ \& $a_8$ & \apm0.03 & \apm0.01 & \apm0.2 &
\apm0.02 & \apm0.000 \\
thresholds    & \apm0.02 & \apm0.05 & \apm0.1 & \apm0.01 & \apm0.003 \\
higher orders   & \apm0.03 & \apm0.04 & \apm0.6 &
\epm{0.15}{0.07} & \epm{0.007}{0.004} \\
higher twists       & \apm0.03 &  -   &  -  &  -  & \apm0.004 \\
\hline
theoretical         & \apm0.07 & \apm0.08 & \apm0.8 &
\epm{0.17}{0.010} &\epm{0.009}{0.006} \\
\hline
\end{tabular}
\end{center}
\caption{Contributions to the errors in the determination of the quantities
$g_A$, $\Delta\Sigma(1)$, $\Delta g(1,1\GeV^2)$, $ a_0(\infty)$ and
$\alpha_s(m_Z)$ from the fits described in the text.
}
\end{table}
\end{center}


\begin{thebibliography}{99}
\baselineskip14pt
\bibitem{EMC} EMC Collaboration, J.~Ashman \etal, \PL\vyp{B206}{1988}{364};
                \NP\vyp{B328}{1989}{1}.
\bibitem{SMCpnew}SMC Collaboration, B.~Adeva \etal, \PL\vyp{B412}{1997}{414}.
\bibitem{SMCdnew} SMC Collaboration,  D.~Adams \etal,
                  \PL\vyp{B396}{1997}{338}.
\bibitem{E143new} E143 Collaboration, K.~Abe \etal, \PL\vyp{B364}{1995}{61}.
\bibitem{E143p} E143 Collaboration, K.~Abe \etal, \PRL\vyp{74}{1995}{346}.
\bibitem{E143d} E143 Collaboration, K.~Abe \etal, \PRL\vyp{75}{1995}{25}.
\bibitem{E142} E142 Collaboration, P.L.~Anthony \etal,
               \PR\vyp{D54}{1996}{6620}.
\bibitem{E154nnew} E154 Collaboration, K.~Abe \etal, \PRL\vyp{79}{1997}{26}.
\bibitem{HERMESn} HERMES Collaboration, K. Ackerstaff \etal,
                  \PL\vyp{B404}{1997}{383}.
\bibitem{altarelliross} G.~Altarelli and G.G.~Ross, \PL\vyp{B212}{1988}{391}.
\bibitem{AL} G.~Altarelli and B.~Lampe, \ZP\vyp{C47}{1990}{315}.
\bibitem{efremov} A.V.~Efremov and O.V.~Teryaev, Dubna preprint
                  E2-88-287 (unpublished).
\bibitem{carlitz}R.D.~Carlitz, J.C.~Collins and A.H.~Mueller,
                 \PL\vyp{B214}{1988}{229}.
\bibitem{instantons} S.~Forte, \PL\vyp{B224}{1989}{189};
                     \NP\vyp{B331}{1990}{1};\\
                     G.M.~Shore and G.~Veneziano, \PL\vyp{B244}{1990}{75};
                     \NP\vyp{B381}{1992}{23};\\
                     R.D.~Ball, \PL\vyp{B266}{1991}{473}.
\bibitem{BEK} S.J.~Brodsky, J.~Ellis and , M.~Karliner,
   \PL\vyp{206B}{1988}{309}

\bibitem{NLO} R.~Mertig and W.~L.~van~Neerven, \ZP\vyp{C70}{1996}{637};\\
              W.~Vogelsang, \PR\vyp{D54}{1996}{2023}, \NP\vyp{B475}{1996}{47}.
\bibitem{NNNLO} S.G.~Gorishny and S.~A.~Larin, \PL\vyp{B172}{1986}{109};\\
                S.A.~Larin and J.~A.~M.~Vermaseren, \PL\vyp{B259}{1991}{345}.
\bibitem{AFR} G.~Altarelli, S.~Forte and G.~Ridolfi, in preparation.
\bibitem{LeaderAnselmino} E.~Leader and M.~Anselmino, \ZP\vyp{C41}{1988}{239}.
\bibitem{JaffeManohar} R.L.~Jaffe and A.~Manohar, \NP\vyp{B337}{1990}{509}.
\bibitem{ABFR} G.~Altarelli, R.D.~Ball, S.~Forte and G.~Ridolfi,
               \NP\vyp{B496}{1997}{337}.
\bibitem{oldSLAC}E80 Collaboration, M.J.~Alguard \etal,
       \PRL\vyp{37}{1976}{1261}, \PRL\vyp{41}{1978}{70};\\
       SLAC-Yale Collaboration,G.~Baum \etal, \PRL\vyp{45}{1980}{2000};\\
       E130 Collaboration, G.~Baum \etal, \PRL\vyp{51}{1983}{1135}.
\bibitem{Bj} J.D.~Bjorken, \PR\vyp{148}{1966}{1467}.
\bibitem{AP} G.~Altarelli and G.~Parisi, \NP\vyp{B126}{1977}{298}.
\bibitem{BFRa} R.D.~Ball, S.~Forte and G.~Ridolfi, \NP\vyp{B444}{1995}{287}.
\bibitem{BFRb} R.D.~Ball, S.~Forte and G.~Ridolfi, \PL\vyp{B378}{1996}{255}.
\bibitem{closeroberts} F.~E.~Close and R.~G.~Roberts, \PL\vyp{B336}{1994}{257}.
\bibitem{Heimann} R.L.~Heimann, \NP\vyp{B64}{1973}{429}.
\bibitem{EK} J.~Ellis and M.~Karliner, \PL\vyp{B341}{1995}{397};
             talk in ``The Spin Structure of the Nucleon'', Proc. of the 1995
             Erice School of Nucleon Structure, ed.~B. Frois and V.W.~Hughes,
             ({\tt hep-ph/9601280}).
\bibitem{revs} G.~Altarelli, in ``The challenging questions'', Proc.
     of the 1989 Erice School, A.~Zichichi, ed. (Plenum, New York, 1990);\\
     R.D.~Ball, in ``The Spin Structure of the Nucleon'', Proc. of the 1995
     Erice School of Nucleon Structure, B. Frois and V.W. Hughes, ed.
     {\tt hep-ph/9511330};\\
     S.~Forte, in the proceedings of the ``14th International Conference on
     Particles and Nuclei (PANIC96)'' and ``12th International Symposium on
     High Energy Spin Physics (SPIN96)'', {\tt hep-ph/9610238};\\
     G.~Ridolfi, in the proceedings of the ``International Workshop on Deep
     Inelastic Scattering and Related Phenomena (DIS96)'',
     {\tt hep-ph/9610214};\\
     G.M.~Shore, {\tt hep-ph/9710367}, to appear in the Proceedings of QCD'97,
     Montpellier, France.
\bibitem{FBR} S.~Forte, R.D.~Ball and G.~Ridolfi, in the proceedings of
              the ``International Workshop on Deep Inelastic Scattering
              and Related Phenomena (DIS96)'', {\tt hep-ph/9608399}.
\bibitem{DeRuj} A.~De~R\'ujula et al., \PR\vyp{10}{1974}{1649}.
\bibitem{BF} R.D.~Ball and S.~Forte, \PL\vyp{B335}{1994}{77};
             \vyp{B336}{1994}{77}; \APP\vyp{B26}{1995}{2097}.
\bibitem{prise} M.~A.~Ahmed and G.~G.~Ross, \PL\vyp{B56}{1975}{385};\\
                M.~B.~Einhorn and J.~Soffer, \NP\vyp{B74}{1986}{714};\\
                A.~Berera, \PL\vyp{B293}{1992}{445}.
\bibitem{BFKL} L.N.~Lipatov, {\it Sov. J. Nucl. Phys.} {\bf 23} (1977) 338;\\
               E.A.~Kuraev, L.N.~Lipatov and V.S.~Fadin,
               {\it Sov. Phys. JETP} {\bf 45} (1977) 199;\\
               Ya.~Balitskii and L.N.~Lipatov, {\it Sov. J. Nucl. Phys.}
               {\bf 28} (1978) 822.
\bibitem{revHera} R.D.~Ball and A.~DeRoeck, in the proceedings of
        the ``International Workshop on Deep Inelastic Scattering and Related
        Phenomena (DIS96)'', {\tt hep-ph/9609309}, and ref. therein.
\bibitem{KL} R.~Kirschner and L.~Lipatov, \NP\vyp{B213}{1983}{122}.
\bibitem{BER} B.I.~Ermolaev, S.I.~Manaenkov and M.G.~Ryskin,
              \ZP\vyp{C69}{1996}{259};\\
              J.~Bartels, B.~I.~Ermolaev and M.~G.~Ryskin,
              \ZP\vyp{C70}{1996}{273}; {\tt hep-ph/9603204}.
\bibitem{NMC} NMC Collaboration, M.~Arneodo \etal, \PR\vyp{D50}{1994}{R1}.
\bibitem{CCFR} W.G.~Seligman \etal, \PRL\vyp{79}{1997}{1213} and refs therein.
\bibitem{NMCF2} NMC Collaboration, M.~Arneodo \etal, \PL\vyp{B364}{1995}{107}.
\bibitem{SLACR} L.W.~Whitlow et al., \PL\vyp{B250}{1990}{193}.
\bibitem{corrnuc} L.L.~Frankfurt and M.~Strikman,
                  {\it Nucl. Phys.} {\bf A405} (1983) 557;\\
                  J.L.~Friar et al., \PR\vyp{C42}{1990}{2310};\\
                  C.~Ciofi degli Atti et al., \PR\vyp{C48}{1993}{968}.
\bibitem{alf} S.~Bethke, talk at the ``High-energy Physics International
              Euroconference on Quantum Chromodynamics (QCD 96)'',
              {\tt hep-ex/9609014};\\
              G.~Altarelli, talks at the ``NATO Advanced Study Institute
              on Techniques and Concepts of High-Energy Physics''
              and at the ``Cracow International Symposium on Radiative
              Corrections (CRAD 96)'', {\tt hep-ph/9611239};\\
              P.N.~Burrows, talk at the ``Cracow International Symposium
              on Radiative Corrections (CRAD 96)'', {\tt hep-ph/9612007}.
\bibitem{CRb} F.~E.~Close and R.~G.~Roberts, \PL\vyp{B316}{1993}{165}.
\bibitem{soper} D.E.~Soper and J.C.~Collins, CTEQ NOTE 94/01,
                {\tt hep-ph/9411214}.
\bibitem{PDG} Particle Data Group, \PR\vyp{D54}{1996}{1}.
\bibitem{AlRi} G.~Altarelli and G.~Ridolfi, \NPBPS\vyp{39B}{1995}{106}.
\bibitem{EJ} J.~Ellis and R.~L.~Jaffe, \PR\vyp{\bf D9}{1974}{1444}.
\bibitem{E154th} E154 Collaboration, K.~Abe et al., \PL\vyp{B405}{1997}{180}.
\bibitem{Stratmann} M.~Stratmann, {\tt hep-ph/9710379}.
\bibitem{Leader} E.~Leader, A.V.~Sidorov and D.B.~Stamenov,
   {\tt hep-ph/9708335}.
\bibitem{Raedel} G.~R\"adel, these Proceedings.
\bibitem{GRSV} M.~Gl\"uck, E.~Reya, M.~Stratmann and W.~Vogelsang,
   \PR\vyp{D53}{1996}{4775}.
\bibitem{DeRoeck} A.~De Roeck, these Proceedings.
\bibitem{Nassalski} J.~Nassalski, these Proceedings.
\bibitem{Feltesse} J.~Feltesse, F.~Kunne and E.~Mirkes,
    \PL\vyp{B388}{1996}{832}. 
\bibitem{Bravar} A.~Bravar, D.~von Harrach and A.~Kotzinian,
   {\tt hep-ph/9710266}.
\bibitem{Lampe} B.~Lampe and A.~Ruffing, {\tt hep-ph/9703308}.
\bibitem{Soffer} J.~Soffer and J.M.~Virey,  \NP\vyp{B509}{1998}{297}. 
\bibitem{Maul} M.~Maul, A.~Schafer, E.~Mirkes and G.~R\"adel,
   {\tt hep-ph/9710309}.
\bibitem{Vogelsang} W.~Vogelsang, {\tt hep-ph/9710345}. 
\bibitem{Contreras} J.G.~Contreras, A.~de Roeck and M.~Maul,
   {\tt hep-ph/9711418}.
\bibitem{Gehrmann} T.~Gehrmann, {\tt hep-ph/9710501}, {\tt hep-ph/9710508}. 
\bibitem{Mirkes} E~.Mirkes and S.~Willfahrt, {\tt hep-ph/9711434}. 
\bibitem{DeFlorian} D.~de Florian, O.A.~Sampayo and R.~Sassot,
   {\tt hep-ph/9711440}.
\bibitem{DeFlorianVogelsang} D.~de Florian and  W.~Vogelsang,
   {\tt hep-ph/9712273}.

\vskip12pt
\end{thebibliography}
\end{document}